\documentclass{aa}
\usepackage{graphicx,amsmath}

\def\ebv{\mbox{$\rm E(B-V)$}}
\def\ms{\mbox{$\rm M_\odot$}}
\def\ds{\mbox{$\rm d_\odot$}}
\def\dgc{\mbox{$\rm d_{GC}$}}

\begin{document}

\title{Globular cluster system and Milky Way properties revisited}

\author{E. Bica\inst{1} \and C. Bonatto\inst{1} \and B. Barbuy\inst{2} \and S. Ortolani\inst{3}}

\offprints{C. Bonatto} 

\institute{Universidade Federal do Rio Grande do Sul, Instituto de F\'\i sica, CP\,15051, Porto Alegre 
91501-970, RS, Brazil\\
\email{bica@if.ufrgs.br, charles@if.ufrgs.br}
\and
Universidade de S\~ao Paulo, Dept. de Astronomia, Rua do Mat\~ao 1226, S\~ao Paulo 05508-090, Brazil\\
\email{barbuy@astro.iag.usp.br}
\and
Universit\`a di Padova, Dipartimento di Astronomia, Vicolo dell'Osservatorio 2, I-35122 Padova, Italy\\
\email{ortolani@pd.astro.it}
}

\date{Received --; accepted --}

\abstract{}{Updated data of the 153 Galactic globular clusters are used to readdress fundamental 
parameters of the Milky Way, such as the distance of the Sun to the Galactic center, bulge and halo
structural parameters and cluster destruction rates.}
{We build a reduced sample, decontaminated of the clusters younger than 10\,Gyr, those with retrograde 
orbits and/or evidence of relation to dwarf galaxies. The reduced sample contains 116 globular 
clusters that are tested whether formed in the primordial collapse.}
{The 33 metal-rich globular clusters ($\rm [Fe/H]\geq-0.75$) of the reduced sample extend basically to the 
Solar circle and distribute over a region with projected axial-ratios typical of an oblate spheroidal, 
$\rm\Delta x:\Delta y:\Delta z\approx1.0:0.9:0.4$. Those outside this region appear to be related to 
accretion. The 81 metal-poor globular clusters span a nearly spherical region of axial-ratios 
$\approx1.0:1.0:0.8$ extending from the central parts to the outer halo, although several clusters in the 
external region still require detailed studies to unravel their origin as accretion or collapse. A new 
estimate of the Sun's distance to the Galactic center based on symmetries of the spatial distribution 
of 116 globular clusters is provided with an uncertainty considerably smaller than in previous determinations 
using globular clusters, $\rm R_O=7.2\pm0.3\,kpc$.  The metal-rich and metal-poor radial-density distributions 
flatten for $\rm R_{GC}\leq2\,kpc$ and are well represented over the full Galactocentric distance range both 
by a power-law with a core-like term and S\'ersic's law; at large distances they fall off as $\rm\sim R^{-3.9}$.}
{Both metallicity components appear to have a common origin, which is different 
from that of the dark matter halo. Structural similarities of the metal-rich and metal-poor radial 
distributions with the stellar halo are consistent with a scenario where part of the reduced sample 
was formed in the primordial collapse, and part was accreted in an early period of merging. 
This applies to the bulge as well, suggesting an early merger affecting the central parts of the 
Galaxy. The present decontamination procedure is not sensitive to all accretions (especially prograde) 
during the first Gyrs, since the observed radial density profiles still preserve traces of the 
earliest merger(s). We estimate that the present globular cluster population corresponds to 
$\rm\leq23\pm6\%$ of the original one. The fact that the volume-density radial distributions of 
the metal-rich and metal-poor globular clusters of the reduced sample follow both a core-like 
power-law and S\'ersic's law indicates that we are dealing with spheroidal subsystems in all 
scales.}


\titlerunning{Galactic system of globular clusters}

\authorrunning{E. Bica et al.}

\maketitle

\section{Introduction}
\label{intro}

Globular clusters (GCs) are potential witnesses of the formation processes that gave rise to
the Milky Way. Because of their long-lived nature, GCs formed in the initial phases of the
Galaxy may preserve in their structure and spatial distribution information that is essential 
to probe these early conditions. In this sense, derivation of the present-day spatial distribution 
of GCs as well as their physical properties can be used to infer on the Galaxy formation and 
evolution processes and better trace out the geometry of the Galaxy.

Early models suggested that the Galaxy formed as a consequence of a monolithic, dissipative collapse 
of a single massive, nearly-spherical spinning gas cloud (e.g. Eggen, Lynden-Bell \& Sandage 
\cite{ELS62}; Sandage \cite{S90}). Initial conditions of the collapse included low-metallicity gas and 
a nearly free-fall regime. This process should be reflected in the GC population as a homogeneity in 
certain parameters, such as orbital motions and a restricted age range. However, later work presented 
observational evidence that the present-day GC population results not only from the primordial collapse 
but from merging and captures of smaller neighbouring galaxies early in the history of the Galaxy (Searle
\& Zinn \cite{SZ78}; Zinn \cite{Z80}) or in more recent events such as the accretion and disruption of 
the Sagittarius dwarf spheroidal galaxy (Ibata, Irwin \& Gilmore \cite{IIG94}; Ibata et al. \cite{I97}). 
On theoretical grounds, N-body simulations (using standard cosmological conditions such as cold dark matter) 
suggest that the hierarchical merging of satellites might be the main building blocks of galaxy formation 
(Bellazzini, Ferraro \& Ibata \cite{MFI2003}).

Present-day data picture the region interior to $\rm\sim10\,kpc$ (inner halo/bulge) as formed 
essentially by the primordial dissipative collapse (e.g. van den Bergh \& Mackey \cite{MB05}), while the 
region external to $\rm\sim15\,kpc$ (outer halo) was formed by later infall and capture of smaller
fragments (Searle \& Zinn \cite{SZ78}; Zinn \cite{Z93}). Mackey \& Gilmore (\cite{MG04}) 
and Mackey \& van den Bergh (\cite{MB05}) studied the properties of GCs by means of the horizontal-branch 
morphology, photometric and structural parameters. They found significant differences among age/metallicity 
sub-groups. Mackey \& Gilmore (\cite{MG04}) reported that $\sim30\%$ of the Milky Way GCs 
have properties similar to those of the GCs in the LMC, SMC, Fornax and Sagittarius dwarf spheroidal 
galaxies. This suggests that a significant fraction of the Galactic GC population, in particular 
outer halo ones, has an extragalactic origin. 

Regardless of the origin, the Galaxy contains $\sim150$ GCs that are  
characterized by a bimodal metallicity distribution, and distances to the Galactic center of up to $\rm\sim150\,kpc$. 
With respect to the metallicity vs. position relation, the metal-rich GCs used to be associated with a disk 
while the metal-poor ones with the halo (Kinman \cite{K59}; Zinn \cite{Z85}; Armandroff \cite{Arm89}). 
More recently the metal-rich GCs have been found to characterize a bulge population 
(Minniti \cite{Min95}; Barbuy et al. \cite{BOBD99}; Cot\'e \cite{Cote99}; van den Bergh \cite{vdB00}).
Chemical enrichment models of the Galaxy, especially for the central parts (Matteucci \& Romano \cite{MR99};
Matteucci, Romano \& Molaro \cite{MRM99}),
predict that the formation of the bulge occurred from the same gas, but even faster than the inner halo.

Considerable efforts have been undertaken in the last $\sim15$ years to obtain fundamental parameters 
of globular clusters by means of accurate CCD colour-magnitude diagrams e.g. for the central parts 
of the Galaxy (Barbuy, Bica \& Ortolani \cite{BBO98}). Harris (\cite{Harris96}), as 
updated\footnote{at http://phy\-sun.\-phy\-sics.mc\-mas\-ter.\-ca/Glo\-bu\-lar.\-html} in 2003 (and 
references therein), compiled parameters of GCs that we will adopt as the starting point for this 
work. Hereafter we refer to Harris' database as H03.

Previous work have focused on the study of global properties and correlations among intrinsic parameters 
of the Galactic GC system, including as well the search for correlations with position in the Galaxy, e.g.
Djorgovski \& Meylan (\cite{DM94}), van den Bergh (\cite{vdB03}) and Mackey \& van den Bergh (\cite{MB05}). 

The significant improvement in the observational GC data is in itself a reason for checking if classically 
adopted values related to the early formation of the Galaxy and the GC system are still valid. In the 
present study we use an updated set of GC parameters, e.g. reddening, metallicity, distance from the 
Sun, age and orbital motion to address their spatial distribution. Since one of the latest derivations 
of the Galactic center distance using GCs was made by Reid (\cite{Reid93}) we also discuss that value 
now based on the updated GC data. 

Our basic approach is to decontaminate the GC sample of the clusters clearly not related to
the primordial collapse of the Galaxy. We do this by identifying young GCs, those with 
retrograde orbits and/or related to the accretion of dwarf satellite galaxies.

This paper is organized as follows. In Sect.~\ref{updatedGCset} we present the updated GC sample and 
isolate the clusters probably associated to the primordial collapse. In Sect.~\ref{GalStr} we present 
projections of their positions onto the (x,y), (x,z) and (y,z) planes, derive the distance to the Galactic 
center and infer on Galactic structure. In Sect.~\ref{RDGC} we build GC radial density profiles, fit them 
with different analytical functions and discuss GC destruction rates. In Sect.~\ref{Disc} we discuss a 
possible scenario to account for the present spatial distribution of GCs. Concluding remarks are given 
in Sect.~\ref{conclu}.

\section{The updated globular cluster data set}
\label{updatedGCset}

In the last 10 years new entries have been added to the census of Galactic GCs either by means 
of discoveries or identifications of misclassified open clusters, e.g. IC\,1257 (Harris et al. 
\cite{Ha97}), ESO\,280-SC06 (Ortolani, Bica \& Barbuy \cite{OBB2000}), 2MASS-GC01 and GC02 (Hurt 
et al. \cite{Hurt2000}), GLIMPSE-C01 (Kobulnicky et al. \cite{Kob05}), the diffuse cluster-type 
object $\rm SDSSJ1049+5103$ (Willman et al. \cite{WBW2005}), and the recently re-classified GC 
Whiting\,1 (Carraro \cite{Carraro05}). For new and already known GCs H03 has provided constant 
updating of fundamental parameters.

Parameters of the 153 presently known GCs in the Galaxy are listed in Table~\ref{tab1}.
We complemented the H03 database by adding information on new GCs in the last 2 years as 
indicated in the notes to Table~\ref{tab1}. The presently updated GC data set will 
be hereafter referred to as the GC05 sample. 

To estimate errors in distance determinations for the subsequent analyses we took into account 
GC05 and typical distance errors in the literature which are dominated by reddening uncertainties 
(e.g. Barbuy, Bica \& Ortolani \cite{BBO98}). We adopt the following values for the reddening and 
distance error relation: 
$\rm\varepsilon=0.05$ for $\ebv\leq0.2$, $\rm\varepsilon=0.10$ for $0.2<\ebv\leq0.8$,
$\rm\varepsilon=0.15$ for $0.8<\ebv\leq1.5$, $\rm\varepsilon=0.2$ for $1.5<\ebv\leq2.0$, and
$\rm\varepsilon=0.25$ for $\ebv\geq2.0$, where $\rm\varepsilon$ is the fractional error in distance.
Uncertainties in related parameters are obtained by propagation of $\rm\varepsilon$.

As discussed in Sect.~\ref{intro}, the Galactic GC system is not expected to be homogeneous in terms 
of cluster origin. Since we intend to base our analysis on GCs with a high probability of formation 
in the primordial collapse, we exclude from GC05 those with  evidence either of extragalactic origin 
or formation later than the collapse. To this category belong the
GCs with retrograde orbits (e.g. Dinescu, Girard \& van Altena \cite{DGA99}), ages younger than 
10\,Gyr (e.g. Salaris \& Weiss \cite{SW02}), direct relation or tidal debris of dwarf
galaxies (e.g. Forbes, Strader \& Brodie \cite{FSB04}) and finally, luminous GCs with evidence 
of being accreted dwarf galaxy nuclei (e.g. Mackey \& van den Bergh \cite{MB05}). References 
are given in col.~12 of Table~\ref{tab1}. We found that 37 GCs (24\% of GC05) fit into one or more of 
these categories. The remaining 116 GCs are part of what we define hereafter as the reduced sample 
(RS-GC05). One caveat with respect to the decontamination process is that it certainly is not 
sensitive to all events, accretions in particular, dating back to the first Gyrs of the Galaxy. 
For instance, extragalactic GCs accreted in this period with prograde orbits would in the present 
hardly be distinguishable from the Milky Way's native population.

Table~\ref{tab1} contains, by columns: (1) - main GC designation; (2) and (3) - Galactic coordinates; 
(4) - reddening; (5) - metallicity; (6) - distance from the Sun; (7) - input Galactocentric distance; 
(8) - output Galactocentric distance (Sect.~\ref{Ro}); (9)--(11) - input Heliocentric components 
(Sect.~\ref{GalStr}); 
(12) - indicators of non-collapse membership; and (13) - additional GC designations as compiled from
the literature by one of us (E.B.).

\setcounter{table}{0}

\begin{table*}[ht]
\caption[]{Updated Globular Cluster parameters - the GC05 sample}
\begin{tiny}
\label{tab1}
\renewcommand{\tabcolsep}{1.0mm}
\begin{tabular}{lrrccrrrrrrcl}
\hline\hline
&&&&&&&&\multicolumn{3}{c}{Heliocentric}\\
\cline{9-11}
GC&$\ell~~$&b~~&E(B-V)&$\rm[Fe/H]$&\ds&\dgc&\dgc&x~~~~&y~~~~&z~~~~&Notes&Alternative designations\\
  &$(^\circ)$&$(^\circ)$& & &(kpc)&(kpc)&(kpc)&(kpc)&(kpc)&(kpc)&\\
(1)&(2)&(3)&(4)&(5)&(6)&(7)&(8)&(9)&(10)&(11)&(12)&(13)\\
\hline
NGC6723 & 0.07 & -17.30 &  0.05 & -1.12 & 8.8 & 2.6&2.9 & 8.40 & 0.01 & -2.62 & &GCl-106,ESO396SC10\\
NGC6287 & 0.13 & 11.02 &  0.60 & -2.05 & 8.5 & 1.7&2.0 & 8.34 & 0.02 & 1.62 & &GCl-54,ESO518SC10\\
NGC6558 & 0.20 & -6.03 &  0.44 & -1.44 & 7.4 &1.0& 0.8 & 7.36 & 0.03 & -0.78 & &Mel-194,Cr368,OCl-3,GCl-89,ESO456SC62\\
NGC6569 & 0.48 & -6.68 &  0.55 & -0.86 & 8.7 & 1.2&1.7 & 8.64 & 0.07 & -1.01 & &GCl-91,ESO456SC77\\
Pal5 & 0.85 & 45.86 &  0.03 & -1.41 & 23.2 & 18.6&18.9 & 16.16 & 0.24 & 16.65 &$Y^{11}$ &Serpens,GCl-32\\
NGC6325 & 0.97  &  8.00 &  0.89 & -1.17 & 9.6 & 2.0&2.6 & 9.51 & 0.16 & 1.34 & &GCl-58,ESO519SC11 \\
NGC6522 & 1.02 & -3.93 &  0.48 & -1.44 & 7.8 & 0.6 &0.8& 7.78 & 0.14 & -0.53 & &GCl-82,ESO456SC43\\
NGC6528 & 1.14 & -4.17 &  0.54 & -0.04 & 9.1 & 1.3&2.0 & 9.07 & 0.18 & -0.66 & &GCl-84,ESO456SC48\\
NGC6652 & 1.53 & -11.38 &  0.09 & -0.96 & 9.6 & 2.4&2.9 & 9.41 & 0.25 & -1.89 & &GCl-98,ESO395SC11\\
M69 & 1.72 & -10.27 &  0.16 & -0.70 & 8.6 & 1.6&2.0 & 8.46 & 0.25 & -1.53 & &NGC6637,GCl-96,ESO457SC14\\
Pal6 & 2.09  &  1.78 &  1.46 & -1.09 & 7.3 & 0.8&0.4 & 7.29 & 0.27 & 0.23 & &GCl-75,ESO520SC21\\
ESO456SC38 & 2.76 & -2.51 &  0.89 & -0.50 & 6.7 & 1.4&0.7 & 6.69 & 0.32 & -0.29 & &Djorgovski2\\
NGC6624 & 2.79 & -7.91 &  0.28 & -0.44 & 8.0 & 1.2&1.4 & 7.91 & 0.39 & -1.10 & &GCl-93,ESO457SC11\\
M70 & 2.85 & -12.51 &  0.07 & -1.51 & 9.0 & 2.1&2.5 & 8.78 & 0.44 & -1.95 & &NGC6681,GCl-101,ESO458SC3\\
NGC6540 & 3.29 & -3.31 &  0.60 & -1.20 & 3.7 & 4.4&3.5 & 3.69 & 0.21 & -0.21 & &Cr364,OCl-11,ESO456SC53,Djorgovski3\\
M107 & 3.37 & 23.01 &  0.33 & -1.04 & 6.4 & 3.3&2.9 & 5.89 & 0.35 & 2.50 & &NGC6171,GCl-44\\
Tz7 & 3.39 & -20.07 &  0.07 & -0.58 & 23.2 & 16.0&16.6 & 21.75 & 1.29 & -7.96 &$S^{1,9}Y^{3,11}$&ESO397SC14\\
NGC6401 & 3.45  &  3.98 &  0.72 & -0.98 & 7.7 & 0.8&0.8 & 7.67 & 0.46 & 0.53 & &GCl-73,ESO520SC11\\
Tz9 & 3.60 & -1.99 &  1.87 & -2.00 & 7.7 & 0.6&0.7 & 7.68 & 0.48 & -0.27 & &ESO521SC11 \\
Tz5 & 3.81  &  1.67 &  2.15 & 0.00 & 7.6 & 0.7&0.7 & 7.58 & 0.50 & 0.22 & &Tz11,ESO520SC26,ESO520SC27\\
M5 & 3.86 & 46.80 &  0.03 & -1.27 & 7.5 & 6.2&5.9 & 5.12 & 0.35 & 5.47 & &NGC5904,GCl-34 \\
Tz10 & 4.42 & -1.86 &  2.40 & -0.70 & 5.7 & 2.4&1.6 & 5.68 & 0.44 & -0.19 & &ESO521SC16\\
NGC6342 & 4.90  &  9.73 &  0.46 & -0.65 & 8.6 & 1.7&2.0 & 8.45 & 0.72 & 1.45 & &GCl-61,ESO587SC6\\
UKS1 & 5.12  &  0.76 &  3.09 & -0.50 & 8.3 & 0.8&1.3 & 8.27 & 0.74 & 0.11 \\
NGC6553 & 5.25 & -3.02 &  0.63 & -0.21 & 5.6 & 2.5&1.7 & 5.67 & 0.51 & -0.30 & &GCl-88,ESO521SC36\\
M9 & 5.54 & 10.70 &  0.38 & -1.75 & 8.2 & 1.7&1.9 & 8.02 & 0.78 & 1.52 & &NGC6333,GCl-60,ESO587SC5 \\
M54 & 5.61 & -14.09 &  0.15 & -1.58 & 19.6 & 27.2&12.8 & 18.92 & 1.86 & -4.77 &$S^{1,9}DN^2$ &NGC6715,GCl-104,ESO458SC8\\
Tz8 & 5.76 & -24.56 &  0.12 & -2.00 & 26.0 & 19.1&19.7 & 23.53 & 2.37  &  -10.81 & $S^{1,9}$&ESO398SC21 \\
NGC6544 & 5.84 & -2.20 &  0.73 & -1.56 & 2.6 & 5.4&4.6 & 2.58 & 0.26 & -0.10 & &Mel-192,Cr366,OCl-17,GCl-87,ESO521SC28\\
NGC6356 & 6.72 & 10.22 &  0.28 & -0.50 & 15.2 & 7.6&8.3 & 14.86 & 1.75 & 2.70 & &GCl-62,ESO588SC1\\
NGC6440 & 7.73  &  3.80 &  1.07 & -0.34 & 8.4 & 1.3&1.7 & 8.31 & 1.13 & 0.56 & &GCl-77,ESO589SC8 \\
M28 & 7.80 & -5.58 &  0.40 & -1.45 & 5.7 & 2.6&1.9 & 5.62 & 0.77 & -0.55 & &NGC6626,GCl-94,ESO522SC23\\
NGC6638 & 7.90 & -7.15 &  0.40 & -0.99 & 8.4 & 1.6&1.9 & 8.26 & 1.15 & -1.05 & &GCl-95,ESO522SC30\\
Tz12 & 8.36 & -2.10 &  2.06 & -0.50 & 4.8 & 3.4&2.6 & 4.75 & 0.70 & -0.18 & &ESO522SC1 \\
Arp2 & 8.55 & -20.78 & 0.10 & -1.76 & 28.6 & 21.4&22.1 & 26.44 & 3.98  &  -10.15 &$S^{1,9}Y^3$ &GCl-112,ESO460SC6 \\
M55 & 8.80 & -23.27 &  0.08 & -1.81 & 5.4 & 3.8&3.2 & 4.90 & 0.76 & -2.13 & &NGC6809,GCl-113,ESO460SC21\\
2MASS-GC02 & 9.78 & -0.62 &  5.56 & ---  & 4.0 & ---&3.4 & 3.94 & 0.68 & -0.04 \\
NGC6642 & 9.81 & -6.44 &  0.41 & -1.35 & 7.7 & 1.6&1.6 & 7.54 & 1.30 & -0.86 & &Mel-203,Cr381,OCl-29,GCl-97,ESO522SC32\\
M22 & 9.89 & -7.55 &  0.34 & -1.64 & 3.2 & 4.9&4.1 & 3.12 & 0.54 & -0.42 & &NGC6656,GCl-99,ESO523SC4\\
2MASS-GC01 & 10.47  &  0.1 &  6.80 & -1.20 & 3.6 & ---&3.7 & 3.54 & 0.65 & 0.01 \\
NGC6717 & 12.88 & -10.9 &  0.22 & -1.29 & 7.4 & 2.3&2.1 & 7.08 & 1.62 & -1.40 & &Pal9,Cr395,OCl-37,GCl-105,ESO523SC14\\
Pal8 & 14.10 & -6.80 & 0.32 & -0.48 & 12.9 & 5.6&6.3 & 12.42 & 3.12 & -1.53 & &GCl-100,ESO591SC12\\
M10 & 15.14 & 23.08 &  0.28 & -1.52 & 4.4 & 4.6&3.9 & 3.91 & 1.06 & 1.72 & &NGC6254,GCl-49\\
M12 & 15.72 & 26.31 &  0.19 & -1.48 & 4.9 & 4.5&3.9 & 4.23 & 1.19 & 2.17  &  &NGC6218,GCl-45 \\
IC1257 & 16.53 & 15.14 &  0.73 & -1.70 & 25.0 & 17.9&18.5 & 16.13 & 6.87 & 6.53 & &OCl-51\\
NGC6366 & 18.41 & 16.04 &  0.71 & -0.82 & 3.6 & 5.0&4.2 & 3.28 & 1.09 & 0.99 &$Y^{11}$ &GCl-65 \\
Pal15 & 18.87 & 24.30 &  0.40 & -1.90 & 44.6 & 37.9&38.5 & 38.46 & 13.15 & 18.35 & &GCl-50 \\
NGC6517 & 19.23  &  6.76 &  1.08 & -1.37 & 10.8 & 4.3&4.7 & 10.13 & 3.53 & 1.27 & &GCl-81\\
M75 & 20.30 & -25.75 &  0.16 & -1.16 & 18.8 & 12.8&13.3 & 15.88 & 5.87 & -8.17 &$R^5$ &NGC6864,GCl-116,ESO595SC13\\
NGC6539 & 20.80  &  6.78 &  0.97 & -0.66 & 8.4 & 3.1&3.2 & 7.80 & 2.96 & 0.99 & &GCl-85 \\
M14 & 21.32 & 14.81 &  0.60 & -1.39 & 8.9 & 3.9&3.9 & 8.02 & 3.13 & 2.27 & &NGC6402,GCl-72 \\
IC1276 & 21.83  &  5.67 &  1.08 & -0.73 & 5.4 & 3.7&3.0 & 4.99 & 2.00 & 0.53 & &Pal7,GCl-90\\
NGC6712 & 25.35 & -4.32 &  0.45 & -1.01 & 6.9 & 3.5&3.1 & 6.22 & 2.95 & -0.52 & &Mel-215,OCl-72,GCl-103\\
M30 & 27.18 & -46.83 &  0.03 & -2.12 & 8.0 & 7.1&6.8 & 4.87 & 2.50 & -5.83 &$R^5$ &NGC7099,GCl-122,ESO531SC21\\
NGC6535 & 27.18 & 10.44 &  0.34 & -1.80 & 6.7 & 3.9&3.5 & 5.86 & 3.01 & 1.21 & &GCl-83\\
NGC6426 & 28.09 & 16.23 &  0.36 & -2.26 & 20.4 & 14.2&14.8 & 17.28 & 9.22 & 5.70 & &Cr346,OCl-81,GCl-76\\
Pal14 & 28.75 & 42.18 &  0.04 & -1.52 & 73.9 & 69.0&69.4 & 48.01 & 26.34 & 49.62 & &AvdB,Arp1,GCl-38\\
Pal12 & 30.51 & -47.68 &  0.02 & -0.94 & 19.1 & 15.9&16.0 & 11.08 & 6.53  &  -14.12 &$Y^{3,11}ST^9$ &Capricornus,GCl-123,ESO600SC11\\
GLIMPSE-C01 & 31.30 & -0.10 &  4.84 & ---  & 3.1 & 6.8&4.8 & 2.65 & 1.61 & -0.01 \\
Pal11 & 31.81 & -15.58 &  0.35 & -0.39 & 12.9 & 7.8&8.1 & 10.56 & 6.55 & -3.46 & &GCl-114\\
M72 & 35.16 & -32.68 &  0.05 & -1.40 & 17.0 & 12.9&13.1 & 11.70 & 8.24 & -9.18 &$R^5$ &NGC6981,GCl-118 \\
NGC6760 & 36.11 & -3.92 &  0.77 & -0.52 & 7.4 & 4.8&4.6 & 5.96 & 4.35 & -0.51 & &Mel-219,OCl-92,GCl-109\\
NGC6749 & 36.20 & -2.20 &  1.50 & -1.60 & 7.9 & 5.0&4.7 & 6.37 & 4.66 & -0.30 & &Be42,OCl-91,GCl-107\\
NGC5466 & 42.15 & 73.59 &  0.00 & -2.22 & 17.0 & 17.2&17.0 & 3.56 & 3.22 & 16.31 & &GCl-27\\
M3 & 42.21 & 78.71 &  0.01 & -1.57 & 10.4 & 12.2&11.8 & 1.51 & 1.37 & 10.20 & &NGC5272,GCl-25\\
NGC6934 & 52.10 & -18.89 & 0.10 & -1.54 & 17.4 & 14.3&14.4 & 10.11 & 12.99 & -5.63 &$R^{5,12}Y^{11}$ &GCl-117\\
Pal10 & 52.44  &  2.72 &  1.66 & -0.10 & 5.9 & 6.4&5.9 & 3.59 & 4.67 & 0.28 & &GCl-111\\
M2 & 53.38 & -35.78 &  0.06 & -1.62 & 11.5 & 10.4&10.2 & 5.56 & 7.49 & -6.72 & &NGC7089,GCl-121\\
NGC7492 & 53.39 & -63.48 &  0.00 & -1.51 & 25.8 & 24.9&24.9 & 6.87 & 9.25  &  -23.09 &$R^5$ &Mel-242,OCl-111,GCl-125 \\
M71 & 56.74 & -4.56 &  0.25 & -0.73 & 3.9 & 6.7&6.0 & 2.13 & 3.25 & -0.31 & &NGC6838,Mel-226,Cr409,OCl-117,GCl-115\\
M13 & 59.01 & 40.91 & 0.02 & -1.54 & 7.7 & 8.7&8.2 & 3.00 & 4.99 & 5.04 & $R^5$&NGC6205,GCl-44 \\
M56 & 62.66  &  8.34 &  0.20 & -1.94 & 10.1 & 9.7&9.4 & 4.59 & 8.88 & 1.46 &$R^{12}$ &NGC6779,GCl-110 \\
NGC7006 & 63.77 & -19.41 &0.05 & -1.63 & 41.5 & 38.8&39.0 & 17.30 & 35.11  &  -13.79 & $R^5$&GCl-119\\
M15 & 65.01 & -27.31 &  0.10 & -1.62 & 10.3 & 10.4&10.1& 3.87 & 8.30 & -4.73 & &NGC7078,GCl-120 \\
M92 & 68.34 & 34.86 &  0.02 & -2.28 & 8.2 & 9.6&9.1 & 2.48 & 6.25 & 4.69 & &NGC6341,GCl-59\\
NGC6229 & 73.64 & 40.31 &  0.01 & -1.43 & 30.7 & 30.0&30.0 & 6.59 & 22.46 & 19.86 & &GCl-47\\
Pal13 & 87.10 & -42.70 &  0.05 & -1.74 & 26.9 & 27.8&27.6 & 1.00 & 19.74  &  -18.24 & $R^7$&Pegasus,GCl-124\\
Pal1 & 130.07 & 19.03 &  0.15 & -0.60 & 10.9 & 17.0&16.3  &  -6.63 & 7.89 & 3.55 &$Y^3CT^9$ &GCl-6\\
NGC288 & 152.28 & -89.38 &  0.03 & -1.24 & 8.3 & 11.6&11.0 & -0.08 & 0.04 & -8.30 &$R^8$ &GCl-2,ESO474SC37\\
SDSSJ1049+5103$^\dag$ & 155.65 & 55.62 &  0.01$^{**}$ & -1.70$^*$ & 45.0 & 50.0&49.1  &  -23.15 & 10.48 & 37.14 \\
Whiting\,1$^\ddag$ & 161.62 & -60.64 &  0.04 & -1.20 & 45.0 & ---&48.1  &  -20.94 & 6.96  &  -39.22 &$Y^4$ &WHI,B0200-03\\
Pal2 & 170.53 & -9.07 &  1.24 & -1.30 & 27.6 & 35.4&34.7  &  -26.88 & 4.48 & -4.35 & &GCl-7\\
NGC2419 & 180.37 & 25.24 &  0.11 & -2.12 & 84.2 & 91.5&90.8  &  -76.16 & -0.49 & 35.90 &$DN^2$ &GCl-12\\
\end{tabular}
\end{tiny}
\end{table*}

\setcounter{table}{0}

\begin{table*}[ht]
\caption[]{Continued}
\begin{tiny}
\renewcommand{\tabcolsep}{1.35mm}
\begin{tabular}{lrrccrrrrrrcl}
\hline\hline
&&&&&&&&\multicolumn{3}{c}{Heliocentric}\\
\cline{9-11}
GC&$\ell~~$&b~~&E(B-V)&$\rm[Fe/H]$&\ds&\dgc&\dgc&x~~~~&y~~~~&z~~~~&Notes&Alternative designations\\
  &$(^\circ)$&$(^\circ)$& & &(kpc)&(kpc)&(kpc)&(kpc)&(kpc)&(kpc)&\\
(1)&(2)&(3)&(4)&(5)&(6)&(7)&(8)&(9)&(10)&(11)&(12)&(13)\\
\hline
Pal4 & 202.31 & 71.80 &  0.01 & -1.48  &  109.2  &  111.8&111.5  &  -31.55  &  -12.95  &  103.74 &$Y^{11}$ &Ursa Majoris,GCl-17 \\
Eridanus & 218.11 & -41.33 &  0.02 & -1.46 & 90.2 & 95.2&94.6  &  -53.29  &  -41.80  &  -59.57 &$Y^{11}$ &ESO551SC1,C0422-213 \\
M79 & 227.23 & -29.35 &  0.01 & -1.57 & 12.9 & 18.8&18.1  &  -7.64 & -8.25 & -6.32 &$C^9$ &NGC1904,GCl-10,ESO487SC7\\
Pal3 & 240.14 & 41.86 &  0.04 & -1.66 & 92.7 & 95.9&95.6  &  -34.37  &  -59.88 & 61.86 &$Y^{11}$ &SextansC \\
NGC1851 & 244.51 & -35.04 &  0.02 & -1.22 & 12.1 & 16.7&16.1  &  -4.26 & -8.94 & -6.95 &$Y^{11}C^9R^5$ &GCl-9,ESO305SC16\\
NGC2298 & 245.63 & -16.01 &  0.14 & -1.85 & 10.7 & 15.7&15.1  &  -4.24 & -9.37 & -2.95 &$C^9$ &GCl-11,ESO366SC22 \\
NGC4147 & 252.85 & 77.19 &  0.02 & -1.83 & 19.3 & 21.3&21.0 & -1.26 & -4.09 & 18.82 &$R^5ST^9$ &GCl-18\\
AM1 & 258.36 & -48.47 &  0.00 & -1.80  &  121.9  &  123.2&123.1  &  -16.31  &  -79.16  &  -91.26 & &E1,ESO201SC10\\
Pyxis & 261.32  &  7.00 &  0.21 & -1.30 & 39.4 & 41.4 &41.1 &  -5.90  &  -38.66 & 4.80 & &Weinberger3\\
NGC1261 & 270.54 & -52.13 &  0.01 & -1.35 & 16.4 & 18.2&17.9 & 0.09  &  -10.07  &  -12.95 &$Y^{11}$ &GCl-5,ESO155SC11 \\
NGC3201 & 277.23  &  8.64 &  0.23 & -1.58 & 5.2 & 9.0&8.4 & 0.65 & -5.10 & 0.78 &$R^{5,6}$ &GCl-15,ESO263SC26\\
NGC2808 & 282.19 & -11.25 &  0.22 & -1.15 & 9.3 & 11.0&10.5 & 1.93 & -8.92 & -1.81 &$C^9Y^{11}$ &GCl-13,ESO91SC1 \\
E3 & 292.27 & -19.02 &  0.30 & -0.80 & 4.3 & 7.6&6.9 & 1.54 & -3.76 & -1.40 & &ESO37SC1\\
M68 & 299.63 & 36.05 &  0.05 & -2.06 & 10.2 & 10.1&9.9 & 4.08 & -7.17 & 6.00 & &NGC4590,GCl-20,ESO506SC30\\
Ru106 & 300.89 & 11.67 &  0.20 & -1.67 & 21.2 & 18.5&18.6 & 10.66  &  -17.82 & 4.29 &$Y^3$ &OCl-887,ESO218SC10\\
NGC4372 & 300.99 & -9.88 &  0.39 & -2.09 & 5.8 & 7.1&6.6 & 2.94 & -4.90 & -1.00 & &GCl-19,ESO64SC6\\
NGC362 & 301.53 & -46.25 &  0.05 & -1.16 & 8.5 & 9.3&8.9 & 3.07 & -5.01 & -6.14 &$Y^{11}R^{12}$ &GCl-3,ESO51SC13 \\
NGC4833 & 303.61 & -8.01 &  0.32 & -1.80 & 6.0 & 6.9&6.4 & 3.29 & -4.95 & -0.84 & &GCl-21,ESO65SC4\\
47\,Tucanae & 305.90 & -44.89 &  0.04 & -0.76 & 4.5 & 7.4&6.7 & 1.87 & -2.58 & -3.18 & &NGC104,GCl-1,ESO50SC9 \\
IC4499 & 307.35 & -20.47 &  0.23 & -1.60 & 18.9 & 15.7&15.9 & 10.74  &  -14.08 & -6.61 &$Y^3$ &GCl-30,ESO22SC5\\
$\omega$ Centauri & 309.10 & 14.97 &  0.12 & -1.62 & 5.3 & 6.4&5.8 & 3.23 & -3.97 & 1.37 &$DN^2R^{12}$ &NGC5139,GCl-24,ESO270SC11\\
NGC5286 & 311.61 & 10.57 &  0.24 & -1.67 & 11.0 & 8.4&8.3 & 7.18 & -8.08 & 2.02 &$CT^9$ &GCl-26,ESO220SC38\\
NGC6101 & 317.75 & -15.82 &  0.05 & -1.82 & 15.3 & 11.1&11.3 & 10.90 & -9.90 & -4.17 & &GCl-40,ESO69SC4 \\
AM4 & 320.28 & 33.51 &  0.04 & -2.00 & 29.9 & 25.5&25.9 & 19.18  &  -15.93 & 16.51 & &AM1353-265 \\
NGC6362 & 325.55 & -17.57 &  0.09 & -0.95 & 8.1 & 5.3&5.1 & 6.37 & -4.37 & -2.45 & &GCl-66,ESO102SC8 \\
NGC5927 & 326.60  &  4.86 &  0.45 & -0.37 & 7.6 & 4.5&4.3 & 6.32 & -4.17 & 0.64 & &GCl-35,ESO224SC4 \\
NGC5946 & 327.58  &  4.19 &  0.54 & -1.38 & 12.8 & 7.4&7.8 & 10.78 & -6.84 & 0.94 & &IC4550,GCl-36,ESO224SC7 \\
BH176 & 328.41  &  4.34 &  0.77 & -0.13 & 14.5 & 8.8&9.2 & 12.32 & -7.57 & 1.10 &$Y^{10}$ &ESO224SC8\\
Lynga7 & 328.77 & -2.79 &  0.73 & -0.62 & 7.2 & 4.2&3.9 & 6.15 & -3.73 & -0.35 & &OCl-949,BH184,ESO178SC11\\
NGC5694 & 331.06 & 30.36 &  0.09 & -1.86 & 34.7 & 29.1&29.6 & 26.20  &  -14.49 & 17.54 & &GCl-29,ESO512SC10\\
NGC5824 & 332.55 & 22.07 &  0.13 & -1.85 & 32.0 & 25.8&26.4 & 26.32  &  -13.67 & 12.02 & &GCl-31,ESO387SC1 \\
M53 & 332.96 & 79.76 &  0.02 & -1.99 & 18.3 & 18.8&18.6 & 2.90 & -1.48 & 18.01 & &NGC5024,GCl-22 \\
NGC5053 & 335.69 & 78.94 &  0.04 & -2.29 & 16.4 & 16.9&16.7 & 2.87 & -1.30 & 16.10 & &Cr267,OCl-970,GCl-23\\
NGC6752 & 336.50 & -25.63 &  0.04 & -1.56 & 4.0 & 5.2&4.5 & 3.31 & -1.44 & -1.73 & &GCl-108,ESO141SC30\\
NGC5986 & 337.02 & 13.27 &  0.28 & -1.58 & 10.5 & 4.8&5.1 & 9.41 & -3.99 & 2.41 & &GCl-37,ESO329SC18\\
NGC6397 & 338.17 & -11.96 &  0.18 & -1.95 & 2.3 & 6.0&5.2 & 2.09 & -0.84 & -0.48 & &GCl-74,ESO181SC4\\
NGC6352 & 341.42 & -7.17 &  0.21 & -0.70 & 5.7 & 3.3&2.7 & 5.36 & -1.80 & -0.71 &$Y^{11}$ &Mel-170,Cr328,OCl-993,GCl-64,ESO228SC3\\
NGC6584 & 342.14 & -16.41 &  0.10 & -1.49 & 13.4 & 7.0&7.4 & 12.23 & -3.94 & -3.79 & &GCl-92,ESO229SC14 \\
NGC5634 & 342.21 & 49.26 &  0.05 & -1.88 & 25.9 & 21.9&22.1 & 16.09 & -5.16 & 19.62 & &GCl-28\\
NGC6139 & 342.37  &  6.94 &  0.75 & -1.68 & 10.1 & 3.6&4.0 & 9.56 & -3.04 & 1.22 & &GCl-43,ESO331SC4\\
NGC5897 & 342.95 & 30.29 &  0.09 & -1.80 & 12.8 & 7.7&8.0 & 10.57 & -3.24 & 6.46 & &GCl-33,ESO582SC2 \\
Tz3 & 345.08  &  9.19 &  0.72 & -0.73 & 7.5 & 2.4&2.2 & 7.15 & -1.91 & 1.20 & &ESO390SC6 \\
NGC6388 & 345.56 & -6.74 &  0.37 & -0.60 & 11.5 & 4.4&5.0 & 11.06 & -2.85 & -1.35 & &GCl-70,ESO279SC2 \\
ESO280SC6 & 346.90 & -12.57 &  0.07 & -2.00 & 21.7 & 14.5&15.0 & 20.63 & -4.80 & -4.72 \\
NGC6256 & 347.79  &  3.31 &  1.03 & -0.70 & 6.6 & 2.1&1.6 & 6.44 & -1.39 & 0.38 & &BH208,ESO391SC6\\
NGC6496 & 348.02 & -10.01 &  0.15 & -0.64 & 11.5 & 4.3&4.9 & 11.08 & -2.35 & -2.00 & &GCl-80,ESO279SC13\\
NGC6541 & 349.29 & -11.18 &  0.14 & -1.83 & 7.0 & 2.2&1.9 & 6.75 & -1.28 & -1.36 & &GCl-86,ESO280SC4 \\
NGC6380 & 350.18 & -3.42 &  1.17 & -0.50 & 10.7 & 3.2&3.8 & 10.52 & -1.82 & -0.64 & &Ton1,Pis25,GCl-68,BH233,ESO333SC14\\
Ton2 & 350.80 & -3.42 &  1.24 & -0.50 & 8.1 & 1.4&1.6 & 7.98 & -1.29 & -0.48 & &Pis26,GCl-71,BH236,ESO333SC16\\
M4 & 350.97 & 15.97 &  0.36 & -1.20 & 2.2 & 5.9&5.2 & 2.09 & -0.33 & 0.61 & &NGC6121,GCl-41,ESO517SC1\\
ESO452SC11 & 351.91 & 12.1 &  0.49 & -1.50 & 7.8 & 2.0&2.0 & 7.55 & -1.07 & 1.64 & &C1636-283\\
NGC6144 & 351.93 & 15.70 & 0.36 & -1.75 & 10.3 & 3.6&4.1 & 9.82 & -1.39 & 2.79 & &GCl-42,ESO517SC6 \\
M80 & 352.67 & 19.46 & 0.18 & -1.75 & 10.0 & 3.8&4.1 & 9.35 & -1.20 & 3.33 & &NGC6093,GCl-39,ESO516SC11\\
NGC6441 & 353.53 & -5.01 & 0.47 & -0.53 & 11.2 & 3.5&4.2 & 11.09 & -1.26 & -0.98 & &GCl-78,ESO393SC34\\
M62 & 353.58  &  7.32 &  0.47 & -1.29 & 6.9 & 1.7&1.2 & 6.80 & -0.77 & 0.88 & &NGC6266,GCl-51,ESO453SC14\\
Liller1 & 354.84 & -0.16 &  3.06 & 0.22 & 10.5 & 2.6&3.4 & 10.46 & -0.94 & -0.03 \\
NGC6453 & 355.72 & -3.87 &  0.66 & -1.53 & 11.2 & 3.3&4.1 & 11.14 & -0.83 & -0.76 & &GCl-79,ESO393SC36\\
NGC6304 & 355.83  &  5.38 &  0.53 & -0.59 & 6.1 & 2.1&1.4 & 6.06 & -0.44 & 0.57 & &GCl-56,ESO454SC2 \\
Tz4 & 356.02  &  1.31 &  2.35  & -1.60 & 9.1 & 1.3&2.0 & 9.08 & -0.63 & 0.21 & &HP4,ESO454SC7 \\
Tz2 & 356.32  &  2.30 &  1.57 & -0.40 & 8.7 & 0.9&1.6 & 8.68 & -0.56 & 0.35 & &HP3,BH228,ESO454SC29\\
Djorgovski1 & 356.67 & -2.48 &  1.44 & -2.00 & 8.8 & 1.0&1.7 & 8.78 & -0.51 & -0.38\\
M19 & 356.87  &  9.38 &  0.41 & -1.68 & 8.7 & 1.6&2.0 & 8.57 & -0.47 & 1.42 & &NGC6273,GCl-52,ESO518SC7 \\
NGC6316 & 357.18  &  5.76 &  0.51 & -0.55 & 11.0 & 3.2&3.9 & 10.93 & -0.54 & 1.10 & &GCl-57,ESO454SC4\\
HP1 & 357.42  &  2.12 &  0.74 & -1.55 & 7.4 & 0.8&0.5 & 7.39 & -0.33 & 0.27 & &GCl-67,BH229,ESO455SC11 \\
Tz1 & 357.57  &  1.00 &  2.28 & -1.30 & 6.2 & 1.8&1.1 & 6.19 & -0.26 & 0.11 & &HP2,GCl-69,ESO455SC23\\
NGC6293 & 357.62  &  7.83 &  0.41 & -1.92 & 8.8 & 1.4&1.9 & 8.71 & -0.36 & 1.20 & &GCl-55,ESO519SC5\\
NGC6284 & 358.35  &  9.94 &  0.28 & -1.32 & 14.7 & 6.9&7.7 & 14.47 & -0.42 & 2.54 & &GCl-53,ESO518SC9\\
Tz6 & 358.57 & -2.16 &  2.14 & -0.50 & 9.5 & 1.6&2.3 & 9.49 & -0.24 & -0.36 & &HP5,BH249,ESO393SC36,BDSB103\\
NGC6235 & 358.92 & 13.52 &  0.36 & -1.40 & 10.0 & 2.9&3.4 & 9.72 & -0.18 & 2.34 & &GCl-58,ESO519SC11\\
NGC6355 & 359.58  &  5.43 &  0.75 & -1.50 & 7.2 & 1.0&0.7 & 7.17 & -0.05 & 0.68 & &Cr330,OCl-1036,GCl-63,ESO519SC15\\
\hline\hline
\end{tabular}
\end{tiny}
\begin{list}{Table Notes.}
\item S: Sagittarius; C: Canis Major; Y: Younger than 10\,Gyr; R: Retrograde orbit; DN: Dwarf nucleus; ST: 
Sagittarius tidal tail; CT: Canis Major tidal tail.
(1) - da Costa \& Armandroff (\cite{CA1995}); (2) - Mackey \& van den Bergh (\cite{MB05}); 
(3) - Rosenberg et al. (\cite{RSPA99}); (4) - Carraro (\cite{Carraro05});
(5) - van den Bergh (\cite{vdB93}); (6) - Cot\'e et al. (\cite{Cote95});
(7) - Siegel et al. (\cite{SMCT01}); (8) - Guo et al. (\cite{GGAL93}); 
(9) - Forbes, Strader \& Brodie (\cite{FSB04}); (10) - Phelps \& Schick (\cite{PS03}); 
(11) - Salaris \& Weiss (\cite{SW02}); (12) - Dinescu, Girard \& van Altena (\cite{DGA99}).
$(\dag)$ and $(\ddag)$ cluster parameters from Willman,  Blanton \& West et al. (\cite{WBW2005}) and
Carraro (\cite{Carraro05}), respectively. $(*)$: based on the conclusion by Willman, Blanton \& West 
et al. (\cite{WBW2005}) that the cluster is more metal-poor than Pal\,5;
$(**)$: based on Schlegel, Finkbeiner \& Davis (\cite{SFD98}).
Col.~7: Galactocentric distance from H03. Col.8~: this work. 
\end{list}
\end{table*}

\begin{figure} 
\resizebox{\hsize}{!}{\includegraphics{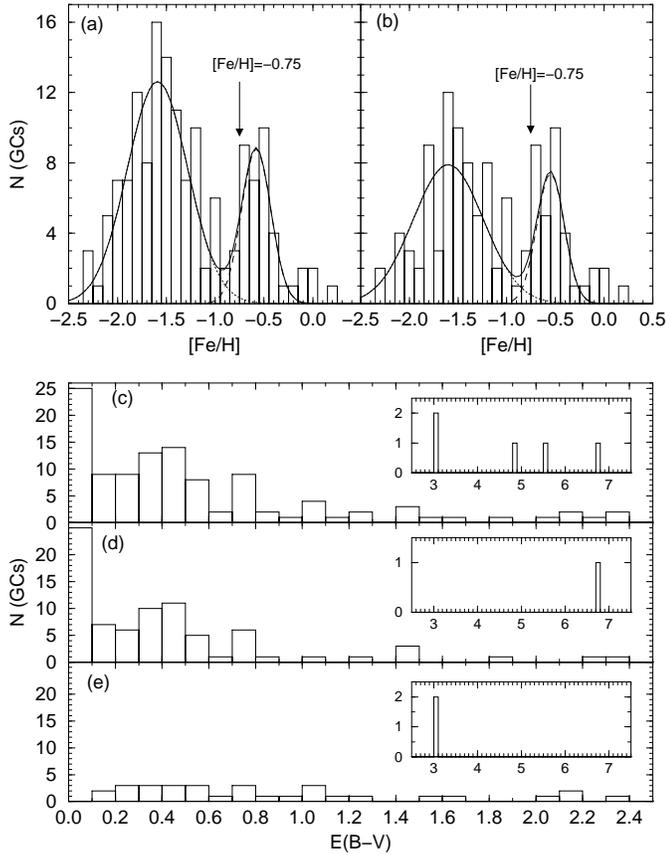}}
\caption[]{GC metallicity distribution of the GC05 (panel (a)) and reduced samples (panel (b)). The
adopted threshold between metal-rich and metal-poor GCs is indicated. The reddening distribution of 
the reduced sample GCs is in panel (c), the corresponding metal-poor ones are in panel (d) and the 
metal-rich ones in panel (e). The insets in panels (c) - (e) show the high-reddening range. 
Note that the two GCs with $\rm\ebv\approx4.8\ and\ 5.5$ (inset of panel c) have no metallicity 
determinations.}
\label{fig1}
\end{figure}

In Fig.~\ref{fig1} (upper panels) we compare the globular clusters of RS-GC05 with those in GC05 in terms of 
metallicity. The well-known bimodal distribution in metallicity (seen e.g. as early as in Zinn \cite{Z85}) 
is confirmed not only in the present GC05 sample (panel (a)) but also in RS-GC05 (panel (b)), however with 
a smaller amplitude ratio between the metal-rich and metal-poor GCs. In the subsequent analysis we adopt 
$\rm [Fe/H]=-0.75$ as the metallicity threshold between metal-rich and metal-poor\footnote{For simplicity 
we refer as metal-poor GCs the genuine together with the intermediate-metallicity ones.} GCs. RS-GC05 
contains 81 metal-poor GCs, 33 metal-rich and 2 with unknown metallicity (2MASS-GC02 and GLIMPSE-C01). 
The reddening distribution of the RS-GC05 GCs (panel (c)) is compared to those of the corresponding metal-poor (panel 
(d)) and metal-rich GCs (panel (e)). The reddening distribution of the metal-poor GCs presents a maximum 
around $\rm\ebv\approx0.05$ and a smaller one at $\rm\ebv\approx0.45$. The low-reddening values are related 
mostly to halo GCs, while the high-reddening ones belong to a more central metal-poor component that spatially 
coexists with the metal-rich bulge clusters (Barbuy, Bica \& Ortolani \cite{BBO98}). The metal-rich GCs, in 
contrast, present a rather uniform distribution in the range $\rm0.05\leq\ebv\leq1.3$.

\section{Galactic structure and the distance to the center}
\label{GalStr}

\begin{figure} 
\resizebox{\hsize}{!}{\includegraphics{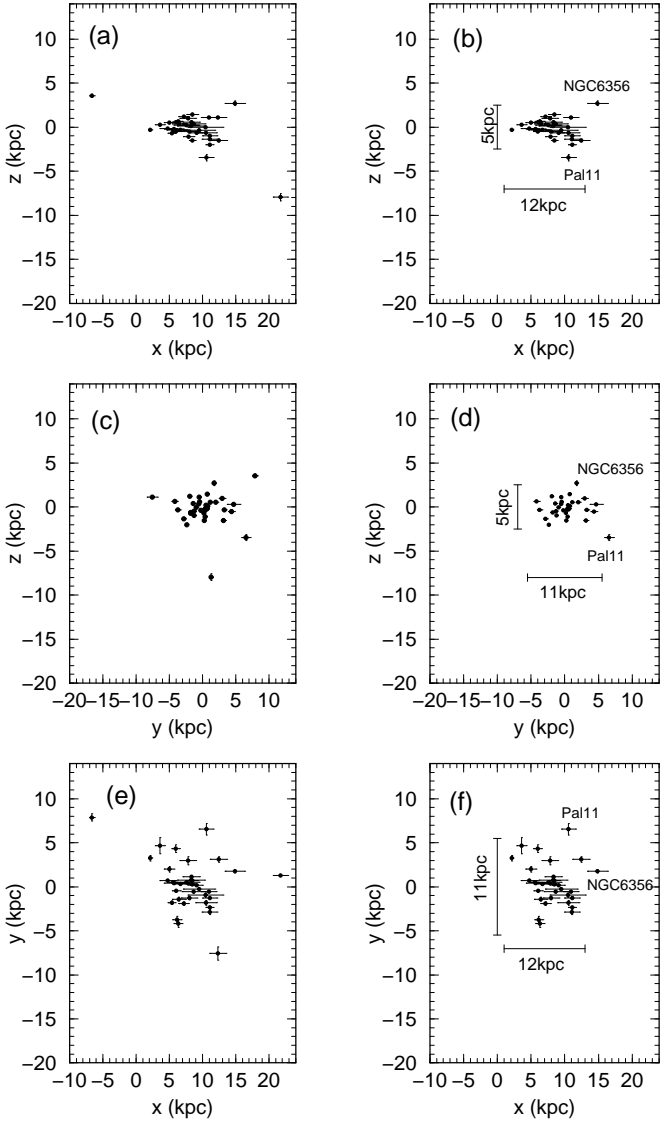}}
\caption[]{Spatial projections of the heliocentric positions of the metal-rich GCs of the 
GC05 (left panels) and reduced (right panels) samples. The reduced sample produces
more concentrated distributions. The outlier metal-rich GCs NGC\,6356 and Palomar\,11 are identified in
panels (b), (d) and (f).}
\label{fig2}
\end{figure}

Inferences on the geometry of the GC system can be made by means of cluster positions 
projected onto the (x,y,z)-heliocentric coordinate axes (Table~\ref{tab1}). In this
coordinate system the x-direction increases from the Sun towards the Galactic center, y is positive 
for $\ell=0^\circ - 180^\circ$ and z increases towards the north Galactic pole. We consider separately 
the metal-rich and metal-poor GCs, both of the GC05 and RS-GC05 samples. In Fig.~\ref{fig2} we show 
the positions of the metal-rich GCs projected onto the (x,y), (x,z) and (y,z) planes for the GC05 
(left panels) and RS-GC05 (right panels) samples. Because the (x,y,z) coordinates are in the heliocentric 
system, the centroids of the y and z distributions coincide with the Galactic center, while that of the 
x-coordinate is shifted from $\rm x=0$ (Sect.~\ref{Ro}). GC05 metal-poor and metal-rich GCs are more 
widely distributed than those in RS-GC05. This effect is minimized in the RS-GC05 plots because most 
of the GC05 outliers belong e.g. to accreted dwarfs or their debris, have young ages and/or retrograde 
orbits (Table~\ref{tab1}). The 2 remaining outliers in the metal-rich RS-GC05 (panels (b), (d) and (f)), 
NGC\,6356 and Palomar\,11, basically define the outer limits of the metal-rich system, slightly beyond the 
present determination of the Solar circle (see below) -- 8\,kpc (Reid \cite{Reid93}). These clusters 
deserve further attention to clarify whether they are young GCs, thus not related to the collapse,
and/or located in the apogalacticon of their orbits. 

From the plots involving the RS-GC05 sample (Fig.~\ref{fig2}) we estimate that the metal-rich GCs distribute 
essentially in a region with dimensions $\rm\Delta x\simeq12\,kpc$, $\rm\Delta y\simeq11\,kpc$ and $\rm\Delta 
z\simeq5\,kpc$, corresponding to axial-ratios $\rm\Delta x:\Delta y:\Delta z\approx1.0:0.9:0.4$. These 
axial-ratios can be accounted for by an oblate spheroid with $\rm\sim5.5\,kpc$ in radius and $\rm\sim2.5\,kpc$ 
in height, a structure spatially coincident with the bulge. As compared to earlier studies the present x and y 
distributions of the metal-rich GCs are similar in extent (Fig.~\ref{fig2}) because of the minimization of 
observational errors achieved with present-day data. 

\begin{figure} 
\resizebox{\hsize}{!}{\includegraphics{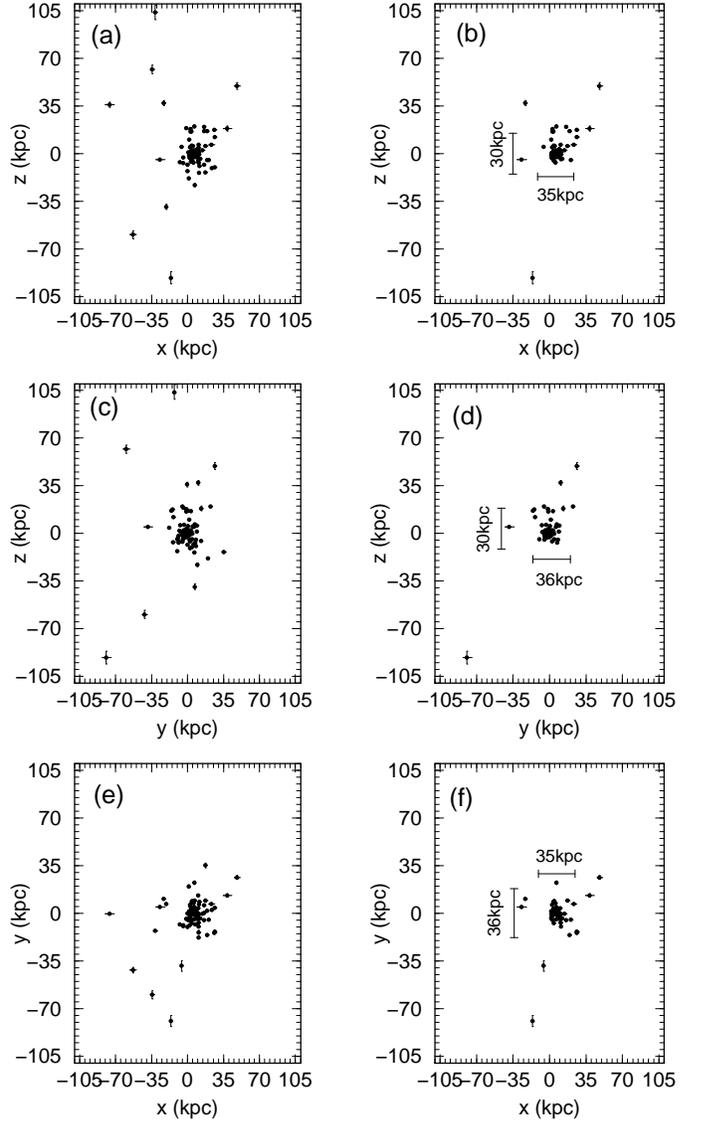}}
\caption[]{Same as Fig.~\ref{fig2} for the metal-poor GCs.}
\label{fig3}
\end{figure}

The metal-poor GCs of RS-GC05 are essentially contained in a region with dimensions 
$\rm\Delta x\simeq35\,kpc$, $\rm\Delta y\simeq36\,kpc$ and $\rm\Delta z\simeq30\,kpc$, with
axial-ratios $\rm\approx1.0:1.0:0.8$. Considering uncertainties these axial ratios 
describe a slightly flattened sphere that reaches into the outer halo.

\begin{figure} 
\resizebox{\hsize}{!}{\includegraphics{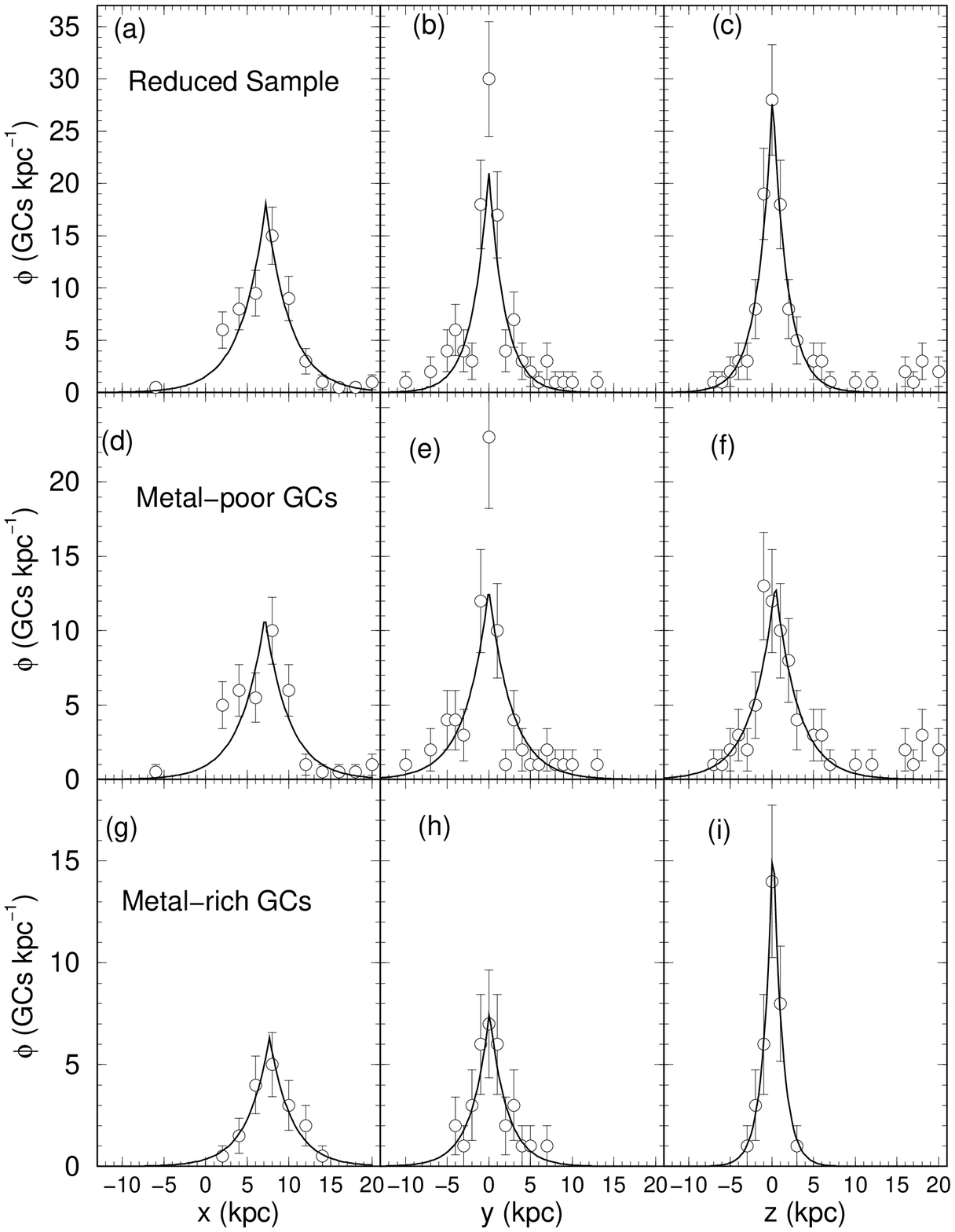}}
\caption[]{One dimensional distribution functions of the reduced sample (top panels), 
metal-poor (middle panels) and metal-rich GCs (bottom panels). The profiles were fitted with 
the exponential-decay function $\rm\phi(\xi)\propto e^{-\left|\frac{\xi-\xi_o}{\xi_h}\right|}$.}
\label{fig4}
\end{figure}

We also infer on the spatial distribution of the GCs in RS-GC05 by means of the distribution function 
$\rm\phi(\xi)=\dfrac{dN}{d\xi}$, which counts the number of GCs in bins of $\rm\Delta\xi=1\,kpc$, 
for the x, y and z coordinates. According to the definition, $\rm\phi(\xi)$ is related to 
the projected one-dimensional number-density of GCs along a given direction. Fig.~\ref{fig4} shows 
the distribution functions of all GCs in RS-GC05, and 
the corresponding metal-poor and metal-rich ones, separately. 
As expected from Figs.~\ref{fig2} and \ref{fig3}, the distribution functions in y and z are 
symmetrical with respect to the centroid of the coordinate system (the Galactic center), while the shift
in x provides the distance of the Sun to the Galactic center (Sect.~\ref{Ro}). The distribution functions
in Fig.~\ref{fig4} can be fitted both with exponential-decay and squared-hyperbolic secant functions. 
Exponential-decay functions usually describe projected surface-density profiles in spiral galaxy disks 
(Binney \& Tremaine \cite{BinTre1987}), while self-gravitating isothermal models such as the squared-hyperbolic 
secant have been used in edge-on disks (e.g. Rice et al. \cite{Rice96}) and lenticular galaxies (van der 
Kruit \& Searle \cite{vdKS81}). Our purpose in fitting the distribution 
functions with a symmetrical profile is to derive the distance of the Sun to the Galactic center 
(Sect.~\ref{Ro}). In this sense we adopted as fit the exponential-decay 
$\rm\phi(\xi)\propto e^{-\left|\frac{\xi-\xi_o}{\xi_h}\right|}$ function, since the respective 
correlation coefficients resulted larger than with squared-hyperbolic secant functions. The fits are shown in 
Fig.~\ref{fig4}, and the resulting central positions ($\xi_o$) and scale-lengths ($\xi_h$) are given in 
Table~\ref{tab2}. The distribution peaks in y and z occur, within uncertainties, at 
$\rm y=z=0$ (Table~2 and Fig.~\ref{fig4}).

Except for the central point in the y-distributions of the RS-GC05 sample (panel (b)) and corresponding 
metal-poor GCs (panel(e)) the exponential-decay function acceptably fits the observed data for distances 
of up to $\rm\sim\pm7\,kpc$ with respect to the peak, within uncertainties. At such distances we are probing 
not only the bulge but the inner halo as well. The fits preserve the symmetrical character of the observed 
profiles, and do not affect the centroid determination. This suggests that not many GCs remain undetected 
towards the central parts, at least to the point of affecting the distance determination.

The individual fit of an exponential-decay profile to each of the (x,y,z) components does not necessarily 
imply a disk structure for the GC subsystems, since we are dealing with one-dimensional distribution 
functions and not surface density profiles.

\begin{table}
\caption[]{Parameters of the one-dimensional distributions}
\label{tab2}
\renewcommand{\tabcolsep}{4.0mm}
\begin{tabular}{lcccc}
\hline\hline
&\multicolumn{3}{c}{Reduced Sample GCs}\\
\cline{2-4}\\
&All&Metal-Poor&Metal-Rich\\
(1)&(2)&(3)&(4)\\
\hline
$\rm x_o~(kpc)$& $7.2\pm0.3$&$7.1\pm0.5$&$7.7\pm0.3$\\
$\rm x_h~(kpc)$& $2.9\pm0.4$&$2.9\pm0.6$&$2.7\pm0.4$ \\
$\rm y_o~(kpc)$& $0.0\pm0.3$&$0.0\pm0.4$&$0.1\pm0.3$ \\
$\rm y_h~(kpc)$& $1.9\pm0.4$&$2.8\pm0.5$&$2.1\pm0.4$\\
$\rm z_o~(kpc)$& $0.1\pm0.2$&$0.4\pm0.3$&$0.1\pm0.1$ \\
$\rm z_h~(kpc)$& $1.8\pm0.2$&$2.7\pm0.5$&$1.1\pm0.1$ \\
$\rm x_h/y_h$& $1.5\pm0.4$&$1.0\pm0.3$&$1.3\pm0.3$ \\
$\rm x_h/z_h$& $1.6\pm0.3$&$1.1\pm0.3$&$2.4\pm0.4$ \\
$\rm y_h/z_h$& $1.1\pm0.3$&$1.0\pm0.2$&$1.9\pm0.4$ \\
\hline
\end{tabular}
\begin{list}{Table Notes.}
\item Parameters of the function $\rm\phi(\xi)\propto e^{-\left|\frac{\xi-\xi_o}{\xi_h}\right|}$ 
fitted to the $\rm\xi=(x,y,z)$ distribution functions. Col.~(2): All GCs of the reduced sample.
\end{list}
\end{table}

The scale-length ratios (Table~\ref{tab2}) derived from the exponential-decay fits agree, within 
uncertainties, with the axial-ratios estimated from Figs.~\ref{fig2} and \ref{fig3}, both for the 
metal-poor and metal-rich GCs of RS-GC05. 

\subsection{Distance to the Galactic center}
\label{Ro}

The distance of the Sun to the Galactic center ($\rm R_O$) has been a recurrent topic in the literature
since Shapley's attempt in 1918 to derive it with globular clusters that resulted in $\rm R_O=13\,kpc$.
Since then different methods with more accurate data and larger GC samples have been used fot the same
purpose. For instance Frenk \& White (\cite{FW82}) using a sample of 65 metal-poor and 11 metal-rich 
GCs limited in latitude to avoid exceedingly large reddening errors affecting distances derived 
$\rm R_O=6.8\pm0.8\,kpc$.

Reid (\cite{Reid93}) reviewed several estimators to derive $\rm R_O$, among them the available GC 
parameters at that time. Estimates based on those GCs put $\rm R_O$ in the range 6.2--10.1\,kpc.
The average $\rm R_O$ from the GC determinations in Table~2 of Reid (\cite{Reid93}) is $\rm7.9\pm1.4\,kpc$, 
which coincides with his best value of $\rm8.0\pm0.5\,kpc$ considering all methods, e.g. 
calibration by OB stars and \ion{H}{i} and \ion{H}{ii} regions, GCs, RR Lyrae and red giants, among 
others. Since then this value has been widely employed in the literature. However, from his Fig.~3 it 
is clear that the x coordinates of the available GCs suffer from reddening/distance uncertainty effects. 
With the GC samples available at the time Maciel (\cite{Maciel93}) and Rastorguev et al. (\cite{R94}) 
obtained $\rm R_O\approx7.6\,kpc$ and $\rm R_O\approx7.0\,kpc$, respectively.
        
More recently, Eisenhauer et al. (\cite{Eis03}) used VLT spectroscopic observations of the orbit of the 
star S2 around SgrA* (assumed to be at the very center of the Galaxy) to derive 
$\rm R_O=8.0\pm0.4\,kpc$. In the same work they provide as well the value of $\rm R_O$ based on the 
statistical parallax distance of 106 late-type and 27 early-type stars located in the central 0.5\,pc. 
They obtained $\rm R_O=7.2\pm0.9\,kpc$.
 
One caveat is that the total-to-selective absorption ratio $\rm R_V=A_V/\ebv$ is not expected to be 
uniform. Variations of $\rm R_V$ in different directions throughout the Galaxy can occur (e.g. Sumi 
\cite{Sumi04}; Ducati, Ribeiro \& Rembold \cite{DRR03}). $\rm R_V$ is also affected by the effective 
wavelength shift in the filters owing to metallicity differences and reddening amount (Barbuy,  Bica   
\& Ortolani \cite{BBO98}, and references therein). Detailed analyses of $\rm R_V$ in the directions 
of all GCs would be necessary to minimize $\rm R_V$-related uncertainties. However, to a first 
approximation we assumed the H03 distances in Table~\ref{tab1}. H03 took into account the 
metallicity dependence of the absolute magnitude of the horizontal branch and from their data
it can be inferred that a constant value of $\rm R_V=3.1$ was adopted throughout.

At the $\rm1\sigma$ level the values of $\rm R_O$ provided by the one-dimensional exponential-decay fits
of the metal-poor (panel (d) of Fig.~\ref{fig4}) and metal-rich GCs (panel (g)) are basically the same
(Table~2). In this sense, to increase the statistical significance of the determination we applied the 
fit to the 116 GCs of RS-GC05 (panel (a) of Fig.~\ref{fig4}). We obtained an average value of $\rm
R_O=7.2\pm0.3\,kpc$. This value puts the Sun $\rm\approx0.8\,kpc$ closer to the Galactic center than
either the best one adopted by Reid (\cite{Reid93}) or that derived by Eisenhauer et al. (\cite{Eis03}). 
However, the present value coincides with that of central stars by Eisenhauer et al. (\cite{Eis03}).
 
The present determination is based on a more accurate and numerous GC database, and consequently, the 
uncertainties in the value of $\rm R_O$ are a factor of $\sim3$ smaller than in previous studies using 
GCs (see Table~2 of Reid \cite{Reid93}). We used the new $\rm R_O$ determination to recalculate the 
Galactocentric distances (col.~8 of Table~\ref{tab1}). 

\subsubsection{Variable total-to-selective absorption}
\label{TTS}

Following the analysis of Barbuy, Bica \& Ortolani (\cite{BBO98}) of the central Galaxy, we now explore the effect 
of a varying total-to-selective absorption as a function of metallicity and reddening amount for the whole GC 
system. According to Grebel \& Roberts (\cite{GR95}) we adopted $\rm R_V=3.6$ for the GCs more metal-rich than 
solar metallicity, and $\rm R_V=3.1$ for $\rm [Fe/H]\leq-1.0$. Interpolation is used for intermediate values
of metallicity. To the metallicity-interpolated $\rm R_V$ we add a further correction related to reddening,
$\rm\Delta R_V=0.05\times\ebv$ (Olson \cite{Olson75}). Dependence of distance on varying $\rm R_V$ can
be expressed as $\rm Ro^\prime=Ro\times10^{\frac{(R_V-R_V^\prime)E(B-V)}{5}}$. 

We applied the above corrections to the data in Table~\ref{tab1} for all metal-rich GCs of RS-GC05 individually, 
leading to a smaller value of $\rm Ro=6.6\pm0.5\,kpc$. Applying the same to all metal-poor GCs of RS-GC05 
individually the distance of the Sun to the Galactic center remains essentially the same as before 
$\rm Ro=7.3\pm0.5\,kpc$ (Table~\ref{tab2}). 

Finally, we consider the ensemble of the metal-rich GCs (Fig.~\ref{fig1}) in order to minimize individual 
uncertainties. The average metallicity and reddening are $\rm [Fe/H]\approx-0.55$ and $\rm\ebv\approx0.5$,
providing an average $\rm R_V^\prime=3.35$. For the metal-rich GCs $\rm R_V=3.1$ (Sect.~\ref{Ro}) and 
$\rm Ro=7.7\pm0.5\,kpc$ (Table~\ref{tab2}), and the resulting distance is $\rm Ro^\prime=7.3\pm0.3\,kpc$, thus fully
compatible with $\rm Ro$ derived from the metal-poor GCs. 

Irrespective of the metallicity and reddening law variations for the metal-rich GCs, the present distance of the 
Sun to the Galactic center determination $\rm Ro\approx7.2\,kpc$ is robust, since it depends essentially on the 
larger sample of metal-poor GCs. This is due to the metal-poor GCs being rather insensitive to $\rm R_V$ assumptions 
and the fact that the current accuracy on their distances is significantly improved.

\section{Radial distribution of globular clusters}
\label{RDGC}

The distribution in Galactocentric distance of the GC number-density, $\rm\rho(R)=\dfrac{dN}{dV}=
\dfrac{dN}{4\pi R^2dR}$, is a potential source of information not only on the present-day Galactic 
structure but the formation processes as well. To investigate this we build radial distribution profiles 
for the metal-rich and metal-poor GCs of RS-GC05 separately. Bins in radius of $\rm\Delta R=0.5\,kpc$ 
for $\rm R_{GC}\leq10\,kpc$ are used to better sample the inner regions, while $\rm\Delta R=2\,kpc$ for 
$\rm 10\leq R_{GC}(kpc)\leq20$ and $\rm\Delta R=10\,kpc$ for $\rm R_{GC}\geq20\,kpc$ to avoid undersampling 
with increasing Galactocentric distance. Taken as face value the radial distribution function as
defined above should be applied to spherically symmetric systems, which is not the case of the oblate
geometry of the metal-rich GCs of RS-GC05 (Sect.~\ref{GalStr}). Implications of this difference in
geometry will be discussed in Sect.~\ref{oblate}.

\subsection{Metal-poor GCs}
\label{MPGC}

The radial distribution of the metal-poor GCs of RS-GC05 is shown in panel (a) of Fig.~\ref{fig5}. 
A fraction of 74\% of the 81 metal-poor GCs is located at Galactocentric distances $\rm R_{GC}\leq10\,kpc$, and 
20\% at $\rm R_{GC}\geq15\,kpc$ (outer halo). The distribution falls off smoothly as a rather steep power-law 
$\rm\sim R^{-(3.6\pm0.2)}$ for $\rm R_{GC}\geq3.5\,kpc$. However, it flattens out for smaller Galactocentric 
distances as $\rm\sim R^{-(1.7\pm0.3)}$. At least part of the flattening might be attributed to 
completeness effects in the crowded central region of the Galaxy. However, a near-IR survey with 2MASS 
(Dutra \& Bica \cite{DuBi2000}) did not reveal any new GC in the central region. Two recent GC discoveries with
2MASS are not 
centrally located, since they are at $l\approx10^\circ$ and near the plane, about halfway from the Sun to the 
Galactic center (Hurt et al. \cite{Hurt2000}). Alternatively, the flattening for small $\rm R_{GC}$ may result 
from the cumulative destruction of GCs close to the Galactic center over a Hubble time (a process that certainly 
played a major r\^ole in depleting the original GC population - see Sect.~\ref{GCDR}), or it may be an intrinsic 
feature of the radial distribution. Modelling of the spatial distribution of the old halo GCs beginning at the 
primordial collapse with cold baryonic gas and dark matter conditions suggests that the inner flattening may 
result not only from tidal destruction, but may in part be of primordial origin (Parmentier \& Grebel \cite{PG05}).

Because of the flattening at small $\rm R_{GC}$ the simplest fits of the observed radial density profile are 
obtained with analytical functions that contain a core-like term. Following Djorgovski \& Meylan (\cite{DM94}) 
we employ the function $\rm\rho(R)=\rho_o/(1+R/R_C)^{\alpha}$, where $\rm R_C$ is the core-like radius. We will 
refer to this function as the composed power-law. The agreement between fit and observed radial distribution 
along the full Galactocentric distance range is excellent (Fig.~\ref{fig5}, panel (a)). The resulting parameters 
are $\rm\rho_o=3.9\pm2.5\,kpc^{-3}$, $\rm R_C=1.5\pm0.6\,kpc$ and $\rm\alpha=3.9\pm0.3$, with a correlation 
coefficient $\rm CC=0.88$. 

Alternatively, in the inset of panel (a) we fit the metal-poor observed radial profile with S\'ersic's 
(\cite{Ser66}) law, $\rm\rho(R)=a\,e^{-b\left[\left(R/R_C\right)^{(1/n)}-1\right]}$. Since it is rather 
insensitive to variations of $\rm R_C$, we used the same core-like radius as that indicated by the composed 
power-law in order to have less free parameters when fitting S\'ersic's law. The best fit was obtained with 
$\rm n=4.1\pm0.7$ and $\rm CC=0.88$.

\subsection{Metal-rich GCs}
\label{MRGC}

The 33 metal-rich GCs of RS-GC05 are contained 
in the region $\rm0.66\leq R_{GC}(kpc)\leq8.3$ (panel (b)), which shows that a sharp radial cutoff thus occurs 
in the metal-rich distribution near the Solar circle. Metal-rich GCs in GC05 located outside this region appear
to be related to accretion of dwarfs and/or young ages (Table~\ref{tab1}). This contrasts with the metal-poor GCs 
that distribute 
in the range $\rm0.36\leq R_{GC}(kpc)\leq123$. Similarly to the metal-poor GCs, a flattening in the radial 
distribution of the metal-rich GCs with respect to the extrapolation of the large Galactocentric distance 
power-law $\rm\sim R^{-(3.2\pm0.2)}$ occurs for $\rm R_{GC}\leq2\,kpc$. This effect should be expected, since 
there is a  lack of correlation of metallicity and GC luminosity (e.g. Djorgovski \& Meylan \cite{DM94}; van 
den Bergh \cite{vdB03}). Parameters of the composed power-law fit are $\rm\rho_o=6.3\pm5.2\,kpc^{-3}$, 
$\rm R_C=1.2\pm1.0\,kpc$ and $\rm\alpha=3.9\pm1.2$, with $\rm CC=0.88$. S\'ersic's law (inset of panel (b)) 
provides a fit with the exponent $\rm n=2.9\pm1.5$ ($\rm CC=0.87$). In this fit we used 
$\rm R_C=1.2\pm1.0\,kpc$, as indicated by the composed power-law. The observed distribution (panel (b) of 
Fig.~\ref{fig5}) cannot be fitted with an exponential-decay law, which precludes the presence of a disk.

Probably as a consequence of the bulge/halo transition, the flattening in both metal-poor and metal-rich 
radial distributions begin at Galactocentric distances compatible with the dimension of the bulge 
(Sect.~\ref{GalStr}), particularly with the (x,y,z) scale-lengths of the metal-rich GCs (Table~\ref{tab2}).

Despite the marked difference in the radial extent of the metal-rich and metal-poor GC profiles, both  
distributions present similar structural features such as flattening in the central region, core-like 
radius ($\rm R_C=1.2 - 1.5\,kpc$), composed ($\rm\alpha=3.9$) and single power-law slopes ($\rm n=3.2 - 3.5$) 
and S\'ersic's law index ($\rm n=2.9 - 4.1$), within uncertainties. These similarities suggest that most 
of the GCs in both metallicity classes share a common origin. 

\subsection{All GCs of the reduced sample}
\label{GCRS}

\begin{figure} 
\resizebox{\hsize}{!}{\includegraphics{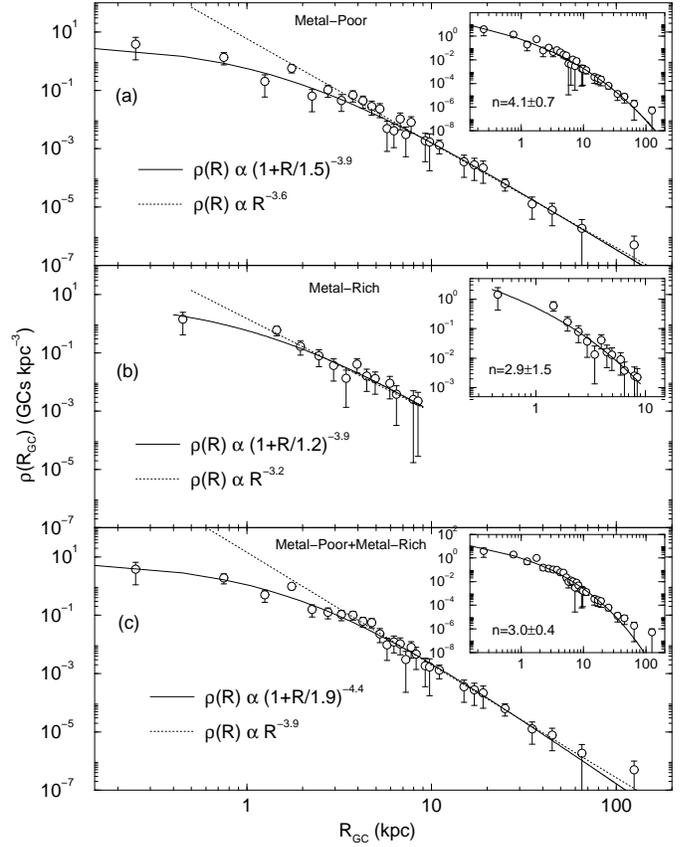}}
\caption[]{Radial density profiles of the GCs in the reduced sample as a function of Galactocentric distance. 
Panel (a) - metal-poor GCs; Panel (b) - metal-rich; Panel (c) - all GCs. Dashed line: single power-law fit for
large Galactocentric distances. Solid line: fit of $\rm\rho(R)=\rho_o/(1+R/R_C)^{\alpha}$. Insets: 
fit of S\'ersic's law $\rm\rho(R)=a\,e^{-b\left[\left(R/R_C\right)^{(1/n)}-1\right]}$.}
\label{fig5}
\end{figure}

The best-fit of the composed power-law to the radial distribution of the 116 RS-GC05 GCs was obtained 
with $\rm\rho_o=7.2\pm3.0\,kpc^{-3}$, $\rm R_C=1.9\pm0.5\,kpc$ and $\rm\alpha=4.4\pm0.3$, with 
$\rm CC=0.91$ (panel (c) of Fig.~\ref{fig5}). Because the metal-rich GCs are contained in the region 
$\rm R_{GC}(kpc)\leq8.3$ the slope of the composed power-law resulted slightly steeper than those of the 
metal-rich and metal-poor distributions, as expected. In addition, the single power-law extrapolation for 
$\rm R_{GC}\geq3.5\,kpc$ falls off as $\rm\sim R^{-(3.9\pm0.1)}$, which within uncertainties basically 
agrees with that of the metal-poor GCs. The best S\'ersic's law fit was obtained with $\rm n=3.0\pm0.4$ 
and $\rm CC=0.89$. Qualitatively, S\'ersic's law and the composed power-law provide essentially the same 
fit to the observed radial profile. 

\subsection{Spherically symmetric-volume densities for the oblate metal-rich sub-system}
\label{oblate}

For practical purposes we assumed spherical symmetry in the above analysis of metal-rich and metal-poor 
radial density profiles. However, the metal-rich GCs distribute through a region whose geometry is 
clearly oblate (Sect.~\ref{GalStr}). Thus, spherically symmetric-volume densities calculated in radial 
bins beyond a few kpc from the center will be artificially decreased, as compared to those measured
in a genuinely (or approximately) spherical system. Consequently, both the measured power-law fall-off 
at large radii and central flattening degree in the metal-rich sub-system might result enhanced relative 
to the nearly spherically symmetric metal-poor sub-system.

\begin{figure} 
\resizebox{\hsize}{!}{\includegraphics{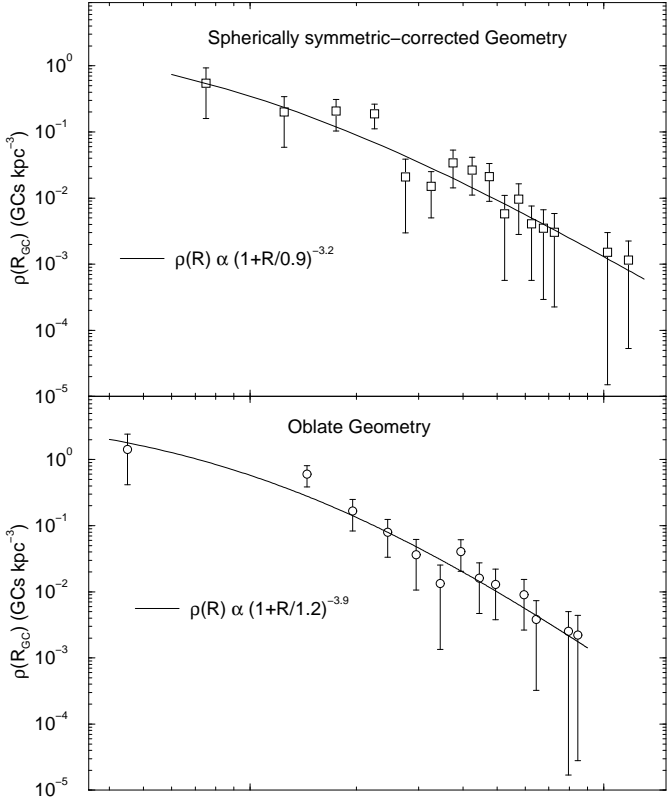}}
\caption[]{Radial density profiles of the metal-rich GCs in the reduced sample measured with 
spherically symmetric volume densities on oblateness-corrected (top panel) and oblate (bottom panel) 
spatial geometries.}
\label{fig6}
\end{figure}

To investigate the effects of non-sphericity in the metal-rich sub-system we apply a coordinate 
transformation to correct for its oblateness, $\rm y\rightarrow y/0.9$ and $\rm z\rightarrow z/0.4$ 
(Sect.~\ref{GalStr}). Subsequently we recalculate Galactocentric distances, 
$\rm R_{GC}=\sqrt{x^2+y^2+z^2}$, and volume densities, $\rm\rho(R)$. Compared to the 
observed oblate profile (bottom panel) the corrected one (top panel of Fig.~\ref{fig6}) presents 
similar shape and somewhat scaled-up Galactocentric distances. Fit of the composed power-law results 
in a core-like $\rm R_C=0.9\pm0.9\,kpc$ and slope $\rm\alpha=3.2\pm0.9$. Both $\rm R_C$ and $\rm\alpha$ 
are similar to the previous ones, but the slope is somewhat flatter, as expected.

Within uncertainties we conclude that measuring spherically symmetric volume densities in the oblate 
metal-rich GC sub-system has a small effect on structural parameters such as power-law fall-off at large
radii and flattening degree in the central region. The effect is minimal probably because of the relatively 
small radial extension ($\rm R_{GC}\sim10\,kpc$) of the metal-rich sub-system. This effect is negligible 
for the nearly-spherical metal-poor GC sub-system (Sect.~\ref{GalStr}).

\subsection{Globular cluster destruction rates}
\label{GCDR}

The discussions in the previous section raised the question whether the flattening observed in the radial 
distribution for regions interior to $\rm R_{GC}\sim2\,kpc$ is a primordial feature or a consequence of enhanced 
GC-destruction rates near the Galactic center. Although the present analysis does not answer this question,
it can be used to provide an estimate on the fraction of primordial GCs still present in the Galaxy. 

The Galactic environment, particularly near the center, tends to destroy star clusters because of enhanced 
tidal truncation and gravitational shocks due to passages close to the bulge and through the disk. The 
bulge, in particular, is very efficient in destroying clusters on highly elongated orbits (Gnedin \& 
Ostriker \cite{GO97}), and the presence of the bar increases the destruction rates by providing a means 
to bring more clusters close to the Galactic center (Long, Ostriker \& Aguilar \cite{LOA92}). Dynamical 
evaporation is probably the most important destruction mechanism in the present-day Galactic environment 
(Gnedin \& Ostriker \cite{GO97}). Aguilar, Hut \& Ostriker (\cite{AHO88}) estimate a current GC depletion 
rate of $\rm\sim5\%$ due basically to dynamical evaporation, indicating that most of the destruction took 
place in the past with bulge-shocking as the main factor. Hut \& Djorgovski (\cite{HD92}) estimate that 
the present GC evaporation rate may be $\rm5\pm3\,Gyr^{-1}$.

Gnedin \& Ostriker (\cite{GO97}) found larger destruction rates than previous work, predicting that 
52\% -- 86\% of the present GC population may be destroyed in the next Hubble time. They conclude 
that the present GC population must be a small fraction of the primordial one, with the debris of 
the destroyed clusters constituting a large fraction of the spheroidal (bulge + halo) stellar population. 
Mackey \& van den Bergh (\cite{MB05}) by means of observational differences in properties of three Galactic 
GC subsystems estimate that the present population could be a fraction $\rm\sim2/3$ of the original one.
Mackey \& Gilmore (\cite{MG04}) estimate a lower limit of 50\% for the destruction rate over the last 
Hubble time, an intermediate value between those of Gnedin \& Ostriker (\cite{GO97}) and Mackey \& 
van den Bergh (\cite{MB05}).

At large Galactocentric distances the efficiency of the GC destruction mechanisms should be minimized 
with respect to the central region. This assumption is supported by numerical simulations showing that 
low-concentration, high-mass GCs are efficiently destroyed in the inner halo but are able to survive at 
$\rm R_{GC}\geq10\,kpc$ (Vesperini \& Heggie \cite{VH97}). For large $\rm R_{GC}$ the number-density of 
GCs in RS-GC05 is well described by the single power-law $\rm\rho(R)=(14.2\pm3.3)R^{-(3.9\pm0.1)}$, the 
discrepancy with respect to the observed profile becoming increasingly larger for $\rm R_{GC}\leq3\,kpc$ 
(Fig.~\ref{fig5} - panel (c)). 

An estimate of the past destruction rate can be derived by comparing the number of GCs in RS-GC05 
with that corresponding to the extrapolation of the large-$\rm R_{GC}$ power-law to the inner regions, 
under the assumption that the difference in both numbers is basically due to the cumulative past 
destruction of GCs. This estimate should be taken as lower-limit since we neglect GC destruction for 
large $\rm R_{GC}$. To derive this value we integrate the large-$\rm R_{GC}$ power-law through the 
Galactocentric distance range over which the 116 GCs of RS-GC05 are observed. Taking into account the 
uncertainties in the parameters of the power-law fit we estimate that the present GC population represents 
a fraction of $\rm\leq23\pm6\%$ of the primordial one. Thus, a lower-limit for the past destruction rate is 
$\sim77\%$ over a Hubble time. This estimate is compatible with that in Gnedin \& Ostriker (\cite{GO97}) 
and about three times larger than that derived by Mackey \& van den Bergh (\cite{MB05}).

The above estimate is based on the assumption that the flattening in the central radial profile is 
essentially due to GC destruction. However, if the flattening is partly primordial, as suggested by
Parmentier \& Grebel (\cite{PG05}), the $\sim77\%$ destruction rate should in fact be taken as an approximate 
upper limit. In this case our estimate would agree with the central range of Gnedin \& Ostriker (\cite{GO97}). 

\subsection{The central kpc}
\label{ckpc}

With the present determination of the Galactocentric distance (Sect.~\ref{Ro}), 9 GCs of the reduced 
sample are located within 1\,kpc of the Galactic center (H03 contains 11 such GCs). The metal-poor 
GCs are Palomar\,6 and HP\,1 at $\rm R_{GC}\leq0.5\,kpc$, and NGC\,6355, Terzan\,9, NGC\,6522, NGC\,6558, 
and NGC\,6401 at $\rm0.5\leq R_{GC}(kpc)\leq1.0$. The metal-rich ones are Terzan\,5 and ESO\,456\,SC38 at 
$\rm0.6\leq R_{GC}(kpc)\leq0.7$. Errors affect these individual estimates, but the ensemble might give 
hints on the extent of a potential avoidance zone, or a central region of enhanced destruction rates 
(e.g. Aguilar, Hut \& Ostriker \cite{AHO88}).

\section{Discussion}
\label{Disc}

In the analysis of the reduced sample GCs we found that the radial density profiles of the
metal-rich and metal-poor GCs are well described by a composed power-law of the form
$\rm\rho(R)\propto(1+R/R_C)^{-\alpha}$, with $\rm3.9\leq\alpha\leq4.4$. For Galactocentric distances 
larger than $\rm\sim2\,kpc$ both observed profiles fall off as a single power-law $\rm R^{-n}$ with 
$\rm3.2\leq n\leq3.9$, while for inner regions they both flatten in a similar way. Structurally, the 
only difference besides geometry (Sect.~\ref{GalStr}) between the spatial distributions of the 
metal-rich and metal-poor GCs is that the former extends basically to the Solar circle, while the latter 
spans from the central parts to the outer halo. This suggests that a significant fraction of the metal-rich 
and metal-poor GCs share a common origin. These conclusions are not significantly affected by the 
oblate and nearly spherical geometries of the metal-rich and metal-poor sub-systems, respectively 
(Sect.~\ref{oblate}).

As pointed out by Djorgovski \& Meylan (\cite{DM94}), the analytical function that we adopted to fit 
the radial density profiles does not have a physically consistent counterpart. However, the steep 
slopes implied by this function for the GC radial density profiles rule out scenarios involving the 
pure monolithic collapse of an isothermal, uniform density cloud. This kind of collapse would produce 
flatter radial density profiles (Abadi, Navarro \& Steinmetz \cite{ANS05}, and references therein). In 
addition, RS-GC05 radial density slopes are significantly steeper than that of the dark matter halo 
(Merritt et al. \cite{MNLJ05}), which in principle precludes a common origin of these structures. 
Consequently, additional mechanisms might have been necessary to increase the density of GCs in the 
central regions of the Milky Way, such as mergers in the early phases.

The rather steep density distribution of the stellar halo was previously derived using globular clusters, 
RR Lyrae stars, blue horizontal-branch (BHB) stars, and star counts. Harris (\cite{Harris76}) and Zinn 
(\cite{Z85}) have shown that the metal-poor GCs distribute radially following a $\rm R^{-n}$ power-law 
profile, with $\rm n=3.5$. Using observations of RR Lyrae Hawkins (\cite{Hawkins84}) derived $\rm n=3.1$ 
with an axial ratio $\rm c/a=0.9$. Bahcall \& Soneira (\cite{BS84}) and Gilmore (\cite{G84}) found 
$\rm c/a\approx0.8$. Using BHB stars Preston, Shectman \& Beers (\cite{PSB91}) found an increase in
the axial ratio from $\rm c/a\approx0.5$ to $\rm c/a\approx1$ up to 20\,kpc with $\rm n=3.5$. Recently, 
Yanny et al. (\cite{Yanny2000}) used BHB tracers from Sloan Digital Sky Survey data to derive $\rm c/a=0.65$ 
and $\rm n=3.2$. Ivezi\'c et al. (\cite{Ivezic2000}) found that the RR Lyrae column density follows a 
shallower power law with $\rm n=2.7$. Robin, Reyl\'e \& Cr\'ez\'e (\cite{Robin2000}) from deep wide field 
star counts estimated a halo flattening of 0.76, and $\rm n=2.44$. 

Evidence of merger events in galaxies has considerably increased in recent years, both on observational 
and theoretical grounds. Deep observations of stars in luminous halos associated with numerical simulations 
of galaxy formation in $\Lambda$-cold dark matter ($\Lambda$CDM) scenarios indicate that a large 
fraction of the stellar content in the halo was not formed {\it in situ}. Abadi, Navarro \& Steinmetz 
(\cite{ANS05} and references therein) suggest that this fraction may have been accreted from 
protogalaxies during earlier merger events. The resulting mass density profiles behave as a power-law 
$\rm R^{-n}$ with $\rm n=3$ at the 
luminous edge of the galaxy and $\rm n\geq4$ more externally, at the virial radius. In addition, the 
density profile of the outer stellar halo is more centrally concentrated with a steeper slope than the 
dark matter halo, whose density profile is caracterized by slopes $\rm n=1-3$ (Merritt et al. 
\cite{MNLJ05}). Hierarchical galaxy formation models under the $\Lambda$CDM framework that are successful 
in reproducing the radial density profile of the Milky Way stellar halo indicate that this structure formed 
from $\sim100$ tidally disrupted, accreted dwarf satellites (Bullock, Kravtsov \& Weinberg \cite{BKW01}). 
What emerges from this is an observational/theoretical picture for the formation and evolution of the stellar 
halo involving violent relaxation and accretion, consistent with hierarchical models of galaxy formation 
(Bellazzini, Ferraro \& Ibata \cite{MFI2003}; Abadi, Navarro \& Steinmetz \cite{ANS05}).

In this sense, the observed similarity of the metal-rich and metal-poor radial density profiles of the 
Galactic GC system with that of the stellar halo suggests that, in addition to the GCs formed in the 
primordial collapse, a non-negligible fraction of the GCs presently in the Milky Way was probably accreted 
during an early period of active merging. This scenario seems to apply to the bulge as well, which raises 
the possiblity of an early merger affecting the central parts of the Galaxy. In the present universe there 
are examples of such mergers, e.g. NGC\,1275 (Zepf et al. \cite{Zepf95}; Holtzman et al. \cite{Holtzman92}). 

Additional support for this scenario comes from the fact that the radial density profiles of the GCs of 
the reduced sample are equally well fitted by S\'ersic's law, 
$\rm\rho(R)\propto e^{-b\left[\left(R/R_C\right)^{(1/n)}-1\right]}$, with $\rm n\approx4.1$ (metal-poor), 
$\rm n\approx2.9$ (metal-rich), and $\rm n\approx3.0$ (combined metal-poor/metal-rich GCs). S\'ersic's 
law with $\rm2\leq n\leq4$ is thought to apply to systems resulting from the mixing that follows from 
violent relaxation or merging (Merritt et al. \cite{MNLJ05}, and references therein). Besides, chemical 
evolution models that reproduce the observed abundances of stars in the bulge suggest that the Galactic 
bulge formed from the same gas but faster than the inner Galactic halo (Matteucci \& Romano \cite{MR99};
Matteucci, Romano \& Molaro \cite{MRM99}). 
The minimum at about $\rm[Fe/H]\approx-0.75$ in the observed metallicity distribution of GCs 
(Fig.~\ref{fig1}) may reflect an external mechanism such as merging to explain the exceedingly 
large number of metal-rich GCs. An early merger in the Milky Way with a relatively massive galaxy might 
have provided the excess metal-rich star formation in the central parts. 

Further evidence on the bulge formation via collapse and/or additional mechanisms will be given by detailed
derivation of metallicities and abundance ratios in comprehensive samples of GCs. However, such detailed
information for bulge GCs is presently scarce. 

\section{Concluding remarks}
\label{conclu}

In this paper the Galactic globular cluster system was decontaminated of the objects with
strong evidence of external origin and/or ages younger than the Galaxy collapse. The resulting 
reduced sample contains 116 GCs, 81 metal-poor and intermediate metallicity clusters ($\rm 
[Fe/H]\leq-0.75$), 33 metal-rich and 2 with unknown metallicity. The classical bimodal metallicity 
distribution is enhanced in the reduced sample.

Projections of the observed heliocentric distances onto the (x,y,z) planes show that the metal-rich 
GCs distribute in a central region of dimensions $\rm\sim12\,kpc\times11\,kpc\times 5\,kpc$, whose 
structure resembles an oblate spheroidal with axial ratio $\rm c/a\approx0.4$. The metal-poor ones 
span a region of dimensions $\rm\sim35\,kpc\times36\,kpc\times30\,kpc$, with a shape similar to a 
slightly flattened sphere with $\rm c/a\approx0.8$. The metal-poor GCs in the reduced sample extend 
into the beginning of the outer halo. Based on the projected number-density of GCs along the x-direction 
we measured the distance of the Sun to the Galactic center as $\rm R_O=7.2\pm0.3\,kpc$. This value 
was obtained considering the spatial distribution of 116 GCs, and is $\rm\sim10\%$ smaller than the 
widely used estimate of Reid (\cite{Reid93}).

Based on structural similarities of the radial density profiles of the present-day GC population with
the stellar halo one can build a scenario where, besides the GCs formed in the primordial collapse, a 
non-negligible fraction of the Milky Way GCs was probably accreted from satellites during an early period 
of merging. Observational and theoretical evidence support this picture, e.g. the GCs formed as 
a consequence of mergers in NGC\,1275 (Zepf et al. \cite{Zepf95}; Holtzman et al. \cite{Holtzman92}).

The present decontamination procedure was not sensitive to all accretions that may have occurred in the 
first Gyrs of the Galaxy, including an eventual major merging, since the observed radial density profiles 
still appear to preserve traces of the earliest merger(s).

Assuming that the flattening in the observed radial density profiles is a consequence of the cumulative
GC depletion mostly by bulge and disk shocking we estimated that the present GC population represents a
fraction of $\rm\leq23\pm6\%$ of the original one. This in turn implies a lower-limit destruction rate of 
$\sim77\%$ over a Hubble time. This estimate is compatible with that of Gnedin \& Ostriker (\cite{GO97}) 
and somewhat larger than that derived by Mackey \& van den Bergh (\cite{MB05}). However, if the
central flattening is partly primordial (Parmentier \& Grebel \cite{PG05}) our estimate would in fact
be an upper limit.
 
The significant improvement in the accuracy of GC data over the last years, as analysed in the
present work, has shed light on the issue whether the metal-rich GCs are associated to a disk (e.g. Zinn 
\cite{Z85}; Armandroff \cite{Arm89}) or a spheroidal subsystem. The fact that the volume-density radial 
distribution of GCs of the reduced sample can be described both by a core-like power-law or a S\'ersic's 
law indicates that the metal-rich GC subsytem is spheroidal. 

The present study pointed out that besides the GC accretions from dwarfs and/or formation later than the
primordial collapse, the radial density distributions require, in addition to a primordial collapse
component a non-negligible early merger population. This scenario provides also a natural explanation
to the second peak in the bimodal metallicity distribution. Through gravitational lensing, large galaxies at 
high-redshift have been detected in the starburst stage, in an epoch compatible to that of the Milky Way's 
primordial collapse. Examples are the $\rm z\sim7$ and $\rm M\sim10^9\,\ms$ galaxy lensed by the 
Abell\,2218 cluster (Egami et al. \cite{Egami05}), and the $\rm z\sim5.5$, $\rm M\sim1-6\times10^{10}\,\ms$ 
starforming galaxy in the field of the cluster RDCS\,1252.9-2927 (Dow-Hygelund et al. \cite{Hygelund05}).
Collapse or its combination with merging are supported by the Galactic GCs and large redshift observations 
of galaxies. On the other hand pure hierarchical galaxy formation has yet to be observed in detail.

\begin{acknowledgements}

E.B., C.B. and B.B. acknowledge support from the Brazilian Institution CNPq. S.O. acknowledges
support from Ministero dell'Universit\`a e della Ricerca Scientifica e Tecnologica (MURST) under 
the program on 'Fasi Iniziali di Evoluzione dell'Alone e del Bulge Galattico' (Italy). We thank
an anonymous referee for comments.
\end{acknowledgements}

%

\end{document}